\documentclass[12pt]{amsart}
\pdfoutput=1

\usepackage{hyperref}

\usepackage{amsmath}
\usepackage[norelsize]{algorithm2e}
\usepackage{scalerel}
\usepackage{tikz-cd}
\usepackage{ulem}
\usepackage{float}
\restylefloat{figure}
\usepackage{stackengine}
\usepackage[english]{babel}
\usepackage[light,condensed,math]{anttor}
\usepackage[T1]{fontenc}
\usepackage{kpfonts}
\usepackage{comment}
\usepackage{array}
\usepackage{lmodern}
\usepackage[mathscr]{euscript}
\usepackage[utf8x]{inputenc}
\usepackage{breqn}
\usepackage{todonotes}
\usepackage{slashed}
\usepackage{youngtab}
\usepackage{amsmath}
\usepackage{amssymb}
\usepackage{amsxtra}
\usepackage{color}
\usepackage[margin=1.2in]{geometry}

\usepackage{float}
\restylefloat{figure}

\newcommand{\boxit}[1]{\vbox{\hrule\hbox{\vrule\kern8pt
\vbox{\hbox{\kern8pt}\hbox{\vbox{#1}}\hbox{\kern8pt}}
\kern8pt\vrule}\hrule}}
\newcommand{\mathboxit}[1]{\vbox{\hrule\hbox{\vrule\kern8pt\vbox{\kern8pt
\hbox{$\displaystyle #1$}\kern8pt}\kern8pt\vrule}\hrule}}

 \newcommand{\beq}{\begin{equation}}
                \newcommand{\bea}{\begin{eqnarray}}
                \newcommand{\eea}{\end{eqnarray}}
                 \newcommand{\eeq}{\end{equation}}

\newcommand {\BC}   {\mathbb C}

\newcommand {\BR}   {\mathbb R}
\newcommand {\BP}   {\mathbb P}

\newcommand {\BT}   {\mathbb T}

\newcommand {\qe} {\mathfrak q}

\newcommand {\ib} {\bar{i}}

\newcommand {\jb} {\bar{j}}

\newcommand {\ii} {\mathrm{i}}

\newcommand {\Det} {\tt Det}

\newcommand {\bx}{  \mathbf{x}}

\newcommand {\bY}{ \mathbf{Y}}

\newcommand {\BS}   {\mathbb S}

\newcommand {\BZ}   {\mathbb Z}

\newcommand {\CalA} {\mathcal A}
\newcommand {\CalB} {\mathcal B}
\newcommand {\CalC} {\mathcal C}

\newcommand {\CalD} {\mathcal D}
\newcommand {\CalE} {\mathcal E}
\newcommand {\CalF} {\mathcal F}

\newcommand {\CalH} {\mathcal H}
\newcommand {\CalI} {\mathcal I}

\newcommand {\CalL} {\mathcal L}

\newcommand {\CalN} {\mathcal N}
\newcommand {\CalO} {\mathcal O}
\newcommand {\CalP} {\mathcal P}
\newcommand {\CalQ} {\mathcal Q}

\newcommand {\CalS} {\mathcal S}
\newcommand {\CalT} {\mathcal T}

\newcommand {\CalX} {\mathcal X}
\newcommand {\CalY} {\mathcal Y}
\newcommand {\CalW} {\mathcal W}
\newcommand {\CalZ} {\mathcal Z}

\newcommand{\ep}{\epsilon}

\renewcommand{\hat}{\widehat}

\newcommand{\Tr}{\mathsf{Tr}\,}

\usepackage{float}
\restylefloat{figure}

\newcommand{\fo}{\vert\kern -.03in\_}

\begin{document}

\title[Tying up instantons with anti-instantons]{Tying up instantons with anti-instantons}

\bigskip
\author[Nikita Nekrasov]{Nikita Nekrasov}

\address{Simons Center for Geometry and Physics\\
Stony Brook University, Stony Brook NY 11794-3636 USA \\
nnekrasov@scgp.stonybrook.edu\footnote{On leave of absence from: IPPI, ITEP, Moscow, Russia}}

\begin{abstract}

In quantizing classical mechanical systems one often sums over the classical trajectories as in localization formulas, but also takes into account the contributions of the "instanton gas": a set of approximate solutions of the equations of motion. This paper attempts to alleviate some of the frustrations of this 40+ years old approach by finding the honest solutions of equations of motion of the complexified classical mechanical system. These ideas originate in the Bethe/gauge correspondence. The examples include algebraic integrable systems, from the abstract Hitchin systems to the well-studied anharmonic oscillator. We also speculate on the applications to the black hole radiation. We elucidate the relation between Lefschetz thimbles and the $\Omega$-deformed $B$-model. We propose the notion of the topological renormalization group.
\bigskip

\bigskip
\centerline{\sl In memory of L.~D.~Faddeev}

\end{abstract}

\maketitle

\vfill\eject

\setcounter{tocdepth}{2}
\tableofcontents

\section{Preface}

I first met L.D.~Faddeev in Leningrad in 1991 during a conference on mathematical physics and, more specifically, quantum groups. I only started my undergraduate studies in Moscow and was very much interested in quantum field theory, string theory and topology.  Moscow and Leningrad schools of theoretical and mathematical physics were competing, so I came from the rivals camp. In fact, one of the physicists at the ITEP with whom I discussed my first undergraduate project: black hole creation in high-energy collisions, told me not to discuss this with L.D. So, understandably, my first impression of L.D. was that of pure fear. I was not alone: all students and almost all professors (I can only think of one person who seemed not to care) who gave talks at the meeting would  look at L.D. in the audience for his approval, fearing his disagreement (and there was some). He was the lion, watching over his pride. His remarks were usually aimed not at the speaker, but at the audience. My last meeting with L.D. was in 2016, in the same city, which by that time became Saint-Petersburg. I lectured at the Chebyshev laboratory on non-perturbative calculations in gauge theory (which we'll discuss below), the lecture was aimed at students, but L.D. nevertheless came by (on his way to meet the President of Russian Federation, to talk him into continuing the support of fundamental research in Russia), and made several remarks, aimed mostly at the audience. 

Twenty five years which passed in between were more or less the second half of L.D.'s active career. His interests in physics and mathematics shaped mine, both through the direct interaction and via the interaction and collaboration with his former students, the members of Faddeev's school. Several of favorite topics of L.D.~ : quantum mechanics, quantum integrable systems, quantum gauge theory, the ``art of quantization'' are now my favorite topics as well. This article is a novel (or so it seemed at the time of writing) approach to the old problem which is at the interface of these themes. 

\section{Introduction}

In the summer of 1995 I was a student at the Ettore Majorana Center summer school on sub-atomic physics in Erice. This place and the school are famous for many reasons. One of them is the lecture course taught by S.~Coleman in 1977 \cite{Col} (see \cite{Pol1, Cal} for the works on which this course was based). I studied the instanton methods using these lecture notes, and of course the book \cite{Pol2}. 

The textbook quantum mechanical problem, which shows the violation of classical intuition, is the non-perturbative splitting of the ground energy level of the quantum mechanical double well potential 
\beq
U(x) = \frac{1}{4} {\lambda} (x^2 - v^2)^2
\label{eq:higgspot}
\eeq
It can be derived using the semi-classical calculation of the tunneling amplitude using the instanton-anti-instanton gas picture. In this picture, the ground state energy is extracted using the small temperature ${\beta} \to \infty$ limit of the matrix elements of the Euclidean evolution operators:
\beq
M_{lr} = \langle - v \vert e^{-{\beta}{\hat H}} \vert +v \rangle\, , \qquad M_{rr} =  \langle + v \vert e^{-{\beta}{\hat H}} \vert +v \rangle
\label{eq:mlrrr}
\eeq
which are represented by the path integrals:
\beq
M_{lr} = \int_{x(0)= v, x({\beta}) = -v} e^{-{\bf S}[x(t)]} Dx(t) \, , \qquad  M_{rr} = \int_{x(0)=  x({\beta}) = v} e^{-{\bf S}[x(t)]} Dx(t)
\label{eq:euclpi}
\eeq
with the Euclidean action
\beq
{\bf S}[x(t)] = \int_{0}^{\beta} \, \frac{{\dot x}^{2}}{2} + U(x)  \ . 
\label{eq:euclac}
\eeq
The critical points of ${\bf S}[x(t)]$ are the classical trajectories $x_{cl}(t)$ in the potential $-U(x)$. For ${\beta} \to \infty$, says the textbook, one has the solution $x_{\rm inst} (t)$ with finite critical value ${\bf S}[x_{\rm inst}(t)] = I$ which describes the particle,  which starts at $x_{\rm inst} (- {\infty}) = +v$ and ends at $x_{\rm inst}(+{\infty}) = -v$, specifically:
\beq
{\dot x}_{\rm inst} = - \sqrt{2U(x_{\rm inst})} = \sqrt{\frac{\lambda}{2}} ( x^2 - v^2 ) \leq 0 \, \Longrightarrow x_{\rm inst} (t) = {\bx} (t_{0}-t) 
\label{eq:inst}
\eeq
where $t_{0}$ is arbitrary, 
\beq
{\bx}(t) = v \, {\rm tanh} \left( {\omega}_{0}t \right)
\label{eq:basinst}
\eeq
and
\beq
2{\omega}_{0} = v \sqrt{2\lambda} = \sqrt{U''(v)}
\eeq
is the frequency of the harmonic oscillator which approximates $U(x)$ near its critical points $x = \pm v$. The critical value 
\beq
I_{0} = {\bf S}[{\bf x}(t)] =  \sqrt{8{\lambda}}\frac{v^{3}}{3}
\eeq
In quantum field theory language $\omega_{0}$ is the mass of the perturbative quanta of the field $x(t)$ in the $0+1$ dimensional theory described by the action 
\beq
S [x(t)] = \int \, dt \, \left( \frac{{\dot x}^{2}}{2} - U(x) \right)
\eeq
in Minkowski time. 

The textbook approach, then, is to note that outside the small interval 
$|t| \leq {\omega}_{0}^{-1}$ the solution can be approximated by the sign function with exponential accuracy:
\beq
{\bx}(t) = v {\rm sign}(t)\, \, ( 1 - e^{-{\omega}_{0}|t|} + \ldots ) \approx  {\bf y}(t) = 
\begin{cases}
- v \, , & \, t < - 2{\omega}_{0}^{-1} \\
 \frac{1}{2} v {\omega}_{0}t \, , & - 2{\omega}_{0}^{-1}  < t < 2 {\omega}_{0}^{-1} \\
+v \, , & \, t >  2{\omega}_{0}^{-1} \\ 
\end{cases}
\label{eq:approx}
\eeq
\bigskip
\noindent
{}and then define the $n$-instanton/antiinstanton (${\CalI}{\bar\CalI}$, in the notations of \cite{Basar:2013eka}) configuration by splicing the trajectory defined by: 
\beq
x (t) = \begin{cases}  {\bf y} ( t_{i}^{-} - t) \, ,  \qquad t_{i-1}^{+} \leq t < t_{i}^{-} \\
 {\bf y} ( t - t_{i}^{+}) \, ,  \qquad t_{i+1}^{-} \geq  t > t_{i}^{+} \\ \end{cases} 
\label{eq:rr}
\eeq
with $t_{0}^{+} = - {\infty}$, 
\beq
t_{1}^{-} \ll t_{1}^{+} \ll t_{2}^{-} \ll t_{2}^{+} \ll \ldots  \ll t_{n}^{-} \ll t_{n}^{+}\, , 
\label{eq:modspa}
\eeq
and $t_{n+1}^{-} = + {\infty}$ (for $M_{rr}$ estimates) and $t_{n}^{-} = + \infty$ (for $M_{lr}$ estimates).

The textbooks then argue, that the contribution of such configurations is, approximately,
\beq
M_{rr,n}({\beta}) = \int dt_{1}^{-} dt_{1}^{+} dt_{2}^{-} dt_{2}^{+} \ldots dt_{n}^{-} dt_{n}^{+} \, K_{rr,n} [x(t)] e^{- {\bf S}[x(t)]} 
\label{eq:mrrn}
\eeq
and
\beq
M_{lr,n} ({\beta})= \int dt_{1}^{-} dt_{1}^{+} dt_{2}^{-} dt_{2}^{+} \ldots dt_{n}^{-}  \, K_{lr,n}[x(t)] e^{- {\bf S}[x(t)]} 
\label{eq:mlrn}
\eeq
where the integrals are taken over the moduli space \eqref{eq:modspa} with lower and upper limits taken to be $\mp {\beta}/2$, with $K$ the  one-loop contribution times the normalization of the ``zero-modes'' $t_{i}^{\pm}$:
\beq
K_{\ldots} [x(t)] = \frac{1}{\sqrt{{\rm Det}' \, \left( {\delta}^{2}{\bf S} \right) }} \sqrt{{\bf g}}
\label{eq:1loop}
\eeq 
With further ingenious approximations one arrives at the estimates $K_{lr, n} \sim K^{2n-1}$, $K_{rr, n} \sim K^{2n}$, making the $n$-${\CalI}{\bar{\CalI}}$ contributions, roughly
\beq
M_{rr,n}({\beta}) \sim \frac{{\beta}^{2n} K^{2n}}{(2n)!} e^{-2n I_{0}} \, , \qquad
M_{lr,n}({\beta}) \sim \frac{{\beta}^{2n-1} K^{2n-1}}{(2n-1)!} e^{-(2n-1) I_{0}} \, , \qquad
\eeq
and, as a result:
\beq
M_{rr} \approx \frac{1}{2} \left( e^{{\beta} K e^{-I_{0}}} + e^{-{\beta} Ke^{-I_{0}}} \right) \, , \qquad
M_{lr} \approx \frac{1}{2} \left( e^{{\beta} K e^{-I_{0}}} - e^{-{\beta} Ke^{-I_{0}}} \right)
\eeq
The trouble is, of course, that the field configurations \eqref{eq:rr} are not, for finite $\beta$, the solutions of ${\delta}{\bf S} = 0$. In fact, expanding ${\bf S}[{\bx}(t) + {\xi}(t)]$ for small ${\xi}(t)$ one finds the non-vanishing tadpoles, i.e. the linear terms in ${\xi}(t)$.  One can tro to remedy this problem by expanding further \cite{B}:
\beq
{\bf S} [{\bx}(t) + {\xi}(t)] = {\bf S} [{\bx}(t)] + \int \, dt\, {\delta}^{(1)}{\bf S} (t) {\xi}(t) + \frac{1}{2} 
\int \, dt_{1} \, dt_{2} {\delta}^{(2)}{\bf S} (t_{1}, t_{2}) {\xi}(t_{1}) {\xi}(t_{2}) + \ldots
\eeq
and then solving for ${\xi}(t)$ in the quadratic approximation, to produce another reference configuration
\beq
{\bx}_{1}(t) = {\bx}(t) - \int\, dt'\, G (t, t')  {\delta}^{(1)}{\bf S} (t') 
\label{eq:bx1}
\eeq
where $G(t', t'')$ is Green's function for the symmetric operator ${\delta}^{(2)}{\bf S}$. According to \cite{B}, this procedure is suggested by A.~Schwarz to be the functional analogue of the well-known Newton's algorithm for finding a true critical point, i.e. the solution to ${\delta}{\bf S} = 0$. 

{}What does this algorithm converges to, if it converges at all? In \cite{Col} it is suggested
that the true critical point of ${\bf S}$ is to be found ``at infinity''. 

\bigskip
\centerline{***}
\bigskip
In this paper we propose another answer to this question, in a class of quantum systems.
This class of systems can be loosely described as the ``quantization of the real slices of the algebraic integrable systems''. An example of such a system is the anharmonic oscillator above. Another example is the many-body elliptic Calogero system:
\beq
{\hat H} = - \frac{\hbar^2}{2} \sum_{i=1}^{N} \frac{{\partial}^2}{{\partial}x_{i}^2} +
{\nu}({\nu}-1) \sum_{i < j} {\wp} (x_{i} - x_{i} ; {\tau})
\eeq
which is the model of $N$ pair-wise repelling particles on a circle of circumference $1$ with the potential having an imaginary (for ${\tau} \in {\ii}{\BR}$) second period. Yet another model would be a spin chain, of XXX, XXZ, XYZ or Gaudin type \footnote{L.D.~ used to approach me with the remark ``You seem to be a nice fellow, but your taste in integrable models is terrible -- here in Leningrad we prefer spin chains to many-body systems ''. Had I taken his advice earlier than I actually had I would have probably missed to connection of integrable systems to gauge theories I will be exploiting below, but would have probably found something else. At any rate, without his work and the work of his school the Bethe/gauge correspondence would not have been discovered. And in the grand picture both many-body systems and the spin chains are on equal footing \cite{FZ, Z}}. 
We shall classify the critical points of the analytically continued action $\bf S$, where the
fields are allowed to take complex values (but the time is kept intact, be it for the Minkowski or Euclidean path integrals).

The paper is organized as follows. In the section $\bf 2$ we review the Bethe/gauge-correspondence. In section $\bf 3$ we describe the supersymmetric gauge theoretic computation of the thermal partition function ${\Tr} e^{-{\beta}{\hat H}}$ of the quantum mechanical system dual to gauge theory on the Bethe side. In section $\bf 4$ we interpret this computation using the topological renormalization group, which treats path integral as a period. In this way we arrive at the problem of finding the solutions of complexified equations of motion, which we identify with the results of the  section $\bf 3$. The reader, not interested in the relations to supersymmetry and gauge theory, can proceed directly to section $\bf 5$ which treats examples. In section $\bf 6$ we summarize and sketch the directions for future research.

{\bf Acknowledgements.} I thank S.~Gukov, I.~Krichever, and E.~Witten for discussions. 
The story presented below has been reported on at several conferences and workshops in the past few years, e.g. "Current Themes in High Energy Physics and Cosmology" at 
the NBI (Copenhagen, Aug 2016)\footnote{https://indico.nbi.ku.dk/event/851/ },  
lectures at the Simons Center for Geometry and Physics (Stony Brook, Sept 2016),  
\footnote{http://scgp.stonybrook.edu/video_portal/video.php?id=2772},\footnote
{http://scgp.stonybrook.edu/video_portal/video.php?id=3028}, Northern California Symplectic Geometry Seminar~/~UC Berkeley String-Math Seminar (Berkeley, Nov 2016), 
Caltech colloquium (Pasadena, Jan 2017), "Hitchin systems in Mathematics and Physics'', Perimeter Institute (Feb 2017) \footnote{https://perimeterinstitute.ca/videos/how-i-learned-stop-worrying-and-love-both-instantons-and-anti-instantons}, 
IGST-2017 in ENS (Paris, Jul 2017)\footnote{https://www.phys.ens.fr/~igst17/}. I thank the organizers of those events for their hospitality and the participants for interesting questions and comments. I would especially like to thank M.~Shaposhnikov for bringing to my attention\footnote{albeit during the presentation of this work at CERN} the works \cite{RR1,RR2,RR3} where the anharmonic oscillator example which we present in some detail below has been worked out for the first time. 

The paper was completed during the visits to the Institut des Hautes {\'E}tudes Scientifiques (Bures-sur-Yvette) and CERN Theory Division. I thank IHES and CERN for their hospitality.

\section{Bethe/gauge-correspondence}

The Bethe/gauge-correspondence connects the gauge theories with ${\CalN}=(2,2)$ $d=2$ super-Poincare invariance (for example, two dimensional gauge theories with four supersymmetries, but also four dimensional ${\CalN}=1$ theories or four dimensional ${\CalN}=2$ theories subject to the two dimensional $\Omega$-deformation) and quantum integrable systems. It identifies the twisted chiral ring (the cohomology of certain nilpotent supercharge ${\CalQ}$) with the set of quantum integrals of motion, the space of supersymmetric vacua with the space of states, the joint eigenvalues of quantum integrals with the vacuum expectation values of the corresponding twisted chiral operators.  In the gauge theory world we don't care much about the reality of these operators. The space of all states of gauge theory is a Hilbert space with the physical Hamiltonian being Hermitian, however the states we are talking about are vacua, i.e. they are annihilated by the Hamiltonian.

Both the first evidence and the main use of this correspondence is the mathematical coincidence of the equations describing the supersymmetric vacua and the equations describing the spectrum of quantum Hamiltonians, in case those are known. 

\subsection{Bethe equations}

In this case they are known as Bethe equations, after H.~Bethe who proposed in 1931 \cite{Bethe1}
an ansatz to describe the eigenstates of the Heisenberg spin chain:
\beq
{\hat H} = \sum_{a=1}^{L} {\vec\sigma}_{a} \otimes {\vec\sigma}_{a+1}
\eeq 
His ansatz reduces the diagonalizing the $2^L \times 2^L$ Hermitian matrix (for the spin chain with $L$ sites, each occupied by a spin $\frac 12$ system)  to solving a system of $N$ algebraic equations on $N$ unknowns ${\lambda}_{1}, \ldots , {\lambda}_{N}$, known now as Bethe roots, or magnon rapidities:
\beq
\prod_{a=1}^{L} \frac{{\lambda}_i - {\mu}_{a} + \frac{\ii}{2}}{{\lambda}_i - {\mu}_{a} - \frac{\ii}{2}} = e^{\ii \vartheta} \, \prod_{j \neq i} \frac{{\lambda}_{i} - {\lambda}_{j} + \ii}{{\lambda}_{i} - {\lambda}_{j} - \ii}  
\label{eq:bae}
\eeq
where $\mu_{1}, \ldots , \mu_L$ are the so-called inhomogeneities of the spin chain (some sort of displacement of the spin sites), and $\vartheta$ describes the quasi-periodic boundary conditions:
\beq
{\vec\sigma}_{a+L} = e^{-\frac{{\ii}{\sigma}_{3} \vartheta}{2}} {\vec\sigma}_{a} e^{\frac{{\ii}{\sigma}_{3} \vartheta}{2}} 
\eeq
The remarkable achievement of Faddeev's school (see \cite{FBA} for the pedagogical introduction) was the invention of the algebraic Bethe ansatz, in which the equations \eqref{eq:bae} ensure that the state
\beq
B({\lambda}_{1})\ldots B({\lambda}_{N}) \vert \Omega \rangle
\eeq
obtained by acting on the (quasi-)vacuum (all spins point down) with the analogues of creation operators is actually the eigenvector of the Hamiltonian and all quantum integrals of motion, generated by the (twisted) transfer-matrix
\beq
A({\lambda}) + e^{{\ii}{\theta}}D({\lambda})
\eeq
The operators $A({\lambda}), B({\lambda}), C({\lambda}), D({\lambda})$ form the so-called Yangian (more precisely the Yangian of $sl(2)$), which is one of the first examples
of quantum groups.

\subsection{Supersymmetric vacua}

The equations \eqref{eq:bae} were recognized \cite{NS1, NS2} as the equations
describing the supersymmetric vacua of a two dimensional supersymmetric gauge theory, 
more specifically a softly broken ${\CalN}=(4,4)$ theory with the gauge group $U(N)$, 
and $L$ fundamental hypermultiplets. The inhomogeneities are the ratios of the masses of these fundamentals to the specific twisted mass responsible for the breaking ${\CalN}=4 \to {\CalN}=2$ (see \cite{NS1} for details), the twist parameter is the (complexified) gauge theta angle (corresponding
to the $U(1)$ factor in the gauge group). 
Equivalently, the equations \eqref{eq:bae} determine the spectrum of the so-called twisted chiral ring, or the $A$-type topological ring. The connection between the topological
gauge theories and quantum integrable (many-body) systems has been observed earlier \cite{GN1, MNS}, and further explored  in \cite{GS1, GS2}, culminating in \cite{NS3}. 

The gauge theory description follows the minimization of the effective potential, 
which due to the low-energy supersymmetry is the norm squared of the derivatives
of the so-called twisted superpotential
${\widetilde{\CalW}}$ so that the vacua are its generalized critical points:
\beq
 \frac{\partial{\widetilde\CalW} ({\sigma})}{{\partial}{\sigma}_{i}} = 2\pi \ii n_{i}\, , \qquad i = 1, \ldots , r
\label{eq:cw}
\eeq
with ${\sigma}_{i}$ being the flat coordinates on the complexified Cartan subalgebra of the gauge group, and $n_i$ integers. Obviously, the equations \eqref{eq:cw} require a special
coordinate system, unlike the more familiar equations describing the vacua of ${\CalN}=2$ Landau-Ginzburg  theories. 
On the quantum integrable side the function $\widetilde{\CalW}$ is identified
with the Yang-Yang potential \cite{YY}, which plays an important role both in the proof of the completeness of Bethe ansatz (where it is complete) and in the understanding the normalizations of Bethe wavefunction \cite{Korepin} and the properties
of the correlation functions \cite{Jimbo:2008kn}.

\subsection{Quantum mechanics from four dimensional gauge theory}

Some quantum integrable systems are sporadic, such as the spin chains of fixed finite spin
at each site. Some come in deformation families, such as the many-body systems. In the latter case the system can be viewed as the quantization of a classical mechanical system (we shall not discuss here to what extent the finite spin systems can be accurately described in terms of the classical mechanics, see the classical work \cite{AFS}). 

We are interested in the systems with variable Planck constant $\hbar$. It turns out that a natural mechanism of getting  such systems within the framework of Bethe/gauge correspondence is to start with ${\CalN}=2$ supersymmetric gauge theory in four dimensions. It can be still viewed as the two dimensional theory, by choosing a $2+1+1$ decomposition ${\BR}^{1,3} \to {\BR}^{1,1} \times {\BR}^{2}$ of the four dimensional spacetime. The rotations of ${\BR}^{2}$ become the two dimensional $SO(2)$ $R$-symmetry. Now turn on the twisted mass for this symmetry \cite{NS3}. This is equivalent to the two-dimensional $\Omega$-deformation of the theory \cite{N2}. More specifically, let us
take the $3+1$ dimensional spacetime to be the product ${\CalD}_{\hbar} \times {\BS}^{1} \times {\BR}^{1} = $ cigar geometry $\times$ a circle $\times$ the time axis. The cigar is the 
two dimensional disk with the rotationally invariant metric $f(r)^2 (dr^2 +  r^2 d{\theta}^2)$, with
$f(r) \to 1$ as $r \to 0$. Let ${\BS}^{1}$ have the circumference $2\pi R_2$. 
The $\Omega$-deformation amounts to replacing the complex adjoint scalar $\phi$ in the vector multiplet
by ${\phi} - {\hbar} D_{\theta}$, so that in particular $D_{A}{\phi} - {\hbar} \iota_{{\partial}_{\theta}} F_{A}$. Note that $\hbar$ is a complex parameter, and
${\phi}^{*} \mapsto {\phi}^{*} - {\hbar}^{*} D_{\theta}$. By choosing $f(r)$ so that $f(r) \to \frac{R_{1}}{r}$ at $r \to \infty$, and then taking $R_{1,2}$ sufficiently small we'll make the four dimensional theory look like 2d sigma model with the worldsheet ${\BR}_{+} \times {\BR}^{1}$ \cite{NW}. The boundary conditions in this sigma model at $r \to 0$ coming from  the absence of singularities in the four dimensional theory happen to be that of the canonical coisotropic brane introduced in \cite{KO} and used in \cite{KW}. Morally speaking 
\cite{GW} the topological sigma model with the canonical coisotropic brane boundary conditions
computes the same three point function as the topological open string introduced in \cite{Ko1} and studied in \cite{CF} (building on the work in \cite{AKSZ}) with the purpose of giving an explicit formula for the  deformation quantization of a Poisson manifold. The concept of deformation quantization is introduced in \cite{Bayen:1977ha}. 

\subsubsection{Many quantum mechanics from four dimensions}
 
 Actually, deformation quantization is not quite a quantization. Deformation quantization produces an associative algebra ${\CalA}_{\hbar}$ over formal (i.e. not necessarily converging) power series over $\hbar$, which in the $\hbar \to 0$ limit becomes  the commutative algebra of functions on the Poisson manifold (the Poisson structure is the first derivative in $\hbar$ of the commutator at $\hbar = 0$). Even if this algebra happens to be well-defined for finite $\hbar$, it need not be representable. However, the construction above produces not only the algebra, it also gives its representation, in fact, a category of representations. These come from the boundary conditions at $r \to \infty$. The full classification is still lacking so we shall be informal.

\subsection{Partition function of the quantum system}

Suppose we do know the supersymmetric gauge theory which is Bethe/gauge dual to our quantum system. 
Using the Bethe/gauge correspondence we can write the following identity for the partition function of the generalized Gibbs ensemble:
\beq
{\CalZ}({\bar\beta}) = {\Tr}_{{\CalH}_{\rm qis}} \, e^{- \frac{1}{\hbar} \sum_{k} {\beta}_{k} {\hat H}_{k}} = {\Tr}_{{\CalH}_{\rm vac}} \, e^{-\frac{1}{\hbar}\sum_{k} {\beta}_{k}  {\CalO}_{k}} \, , 
\label{eq:qmpfq}
\eeq
with ${\bar\beta} = ( \beta_{k} )_{k=1}^{r}$ the set of generalized inverse temperatures, and ${\CalO}_{k}$ the basis of the twisted chiral ring, which we can further express as:
\beq
{\Tr}_{{\CalH}_{\rm qis}} \, e^{- \frac{1}{\hbar}\sum_{k} {\beta}_{k} {\hat H}_{k}} = {\Tr}_{{\CalH}_{\rm vac}} \, e^{-\frac{1}{\hbar}\sum_{k} {\beta}_{k} {\CalO}_{k}} = {\Tr}_{{\CalH}_{\rm vac}} \, (-1)^{F}\, e^{-\frac{1}{\hbar}\sum_{k} {\beta}_{k} {\CalO}_{k}} \,  ,
\label{eq:qmpfq2}
\eeq
assuming all vacua are bosonic,  and as
\begin{multline}
{\Tr}_{{\CalH}_{\rm qis}} \, e^{- \frac{1}{\hbar}\sum_{k} {\beta}_{k} {\hat H}_{k}}  = {\Tr}_{{\CalH}_{\rm vac}} \, (-1)^{F}\, e^{-\frac{1}{\hbar}\sum_{k} {\beta}_{k} {\CalO}_{k}} =  \\
 =  {\Tr}_{{\CalH}_{\rm gauge}} \, (-1)^{F}\, e^{-\frac{1}{\hbar}\sum_{k} {\beta}_{k} {\CalO}_{k}} \, , 
\label{eq:qmpfq3}
\end{multline}
using $[ {\CalQ} , {\CalO}_{k} ] = 0$ and the usual Witten index argument.

Now assume further that the gauge theory in question is the $\Omega$-deformed four dimensional ${\CalN}=2$ gauge theory on ${\BR}^{1,1} \times {\CalD}$, with cigar-type geometry $\CalD$, the $\Omega$-deformation with the parameter $\hbar$ on $\CalD$. 

The corresponding Witten index is now given by the path integral on the four-dimensional 
space-time of the form ${\BT}^{2} \times {\CalD}$, with the $\Omega$-deformation along ${\CalD}$. Now, the twisted chiral ring observables ${\CalO}_{k}$ come from the local observables ${\CalO}_{k}^{(0)}$ in the four dimensional theory (the gauge invariant polynomials of the complex adjoint scalar $\phi$). Their cohomological descendents 
${\CalO}^{(2)}_{k}$ can be integrated over $pt \times \CalD$, with $pt \in {\BT}^{2}$. In the 
$\Omega$-deformed theory
\beq
\frac{1}{\hbar} {\CalO}_{k} (pt \times 0) = \int_{pt \times \CalD} {\CalO}_{k}^{(2)}
\label{eq:loc02}
\eeq
where $0$ is the tip of the cigar. The latter description makes sense even when $\hbar \to 0$. 

Thus, the quantum-mechanical system partition function is equal to the susy partition function of the ${\CalN}=2$ gauge theory on
${\BT}^{2} \times  {\CalD}$. 
\beq
{\CalZ}({\bar\beta}) = \int_{4d\rm  superfields} \, e^{-\int_{{\BT}^{2} \times {\CalD}} \, {\CalL}_{\rm SYM}^{\hbar}} \ \cdot\ e^{\sum_{k} {\beta}_{k} \int_{\CalD} {\CalO}_{k}^{(2)}}
\label{eq:qmpf4}
\eeq
with the perturbation terms ${\beta}_{k} \int_{\CalD} {\CalO}_{k}^{(2)}$ being essentially the Donaldson surface-observables integrated along the cigar $\CalD$, and ${\CalL}_{\rm SYM}^{\hbar}$ the Lagrangian of the $\Omega$-deformed four dimensional ${\CalN}=2$ theory. 

What is implicit in the formula \eqref{eq:qmpf4} is that on the left hand side one has a specific realization of the noncommutative algebra of observables in the (Hilbert) space of states ${\CalH}_{\rm qis}$, while on the right hand side one fixes some supersymmetric
boundary conditions on the infinite end $\partial\CalD$ of the cigar $\CalD$. 

\section{Landau-Ginzburg from gauge theory}

\subsection{Effective superpotential}

Now suppose the boundary conditions on $\partial\CalD$ are matched with ${\CalH}_{\rm qis}$. How can we exploit \eqref{eq:qmpf4}? We can use the knowledge of the low-energy
effective theory to get an estimate of the partition function for small $\hbar$ (and large enough $\beta_k$'s). 

Indeed, let us deform the metric on ${\BT}^{2}$. The usual Witten index arguments guarantee the independence of the partition function on the size of ${\BT}^{2}$. So let us 
make it very large, much larger than ${\Lambda}_{\rm SYM}^{-1}$ scale. Integrate out the high energy modes. We end up with the effective low-energy theory on ${\BT}^{2} \times {\CalD}$. The presence of the compact torus (it could be replaced by any compact Riemann surface at the expense of inserting an additional twisted chiral ring operator under the trace)
implies the existence of the topological sectors in the effective abelian theory, namely the electric and magnetic fluxes through ${\BT}^{2}$:
\beq
m^{i} =  \frac{1}{2\pi \ii} \int_{{\BT}^{2}} F^{(i)}, \qquad n_{i} = \frac{1}{2\pi \ii} \sum_{j=1}^{r} {\tau}_{ij}    \int_{{\BT}^{2}} \star F^{(j)}
\eeq
where $F^{(i)}$ are the abelian gauge curvatures, $i = 1, \ldots , r$, $r$ being the number of the abelian vector multiplets in the effective theory, 
\beq
{\tau}_{ij} = \frac{{\partial}^{2} {\CalF}}{{\partial}a^{i} {\partial} a^{j}}
\label{eq:effcou}
\eeq
being the matrix of effective gauge couplings and theta
angles, and ${\CalF}$ the prepotential of the effective theory. 

Now let us perform the Kaluza-Klein compactification on this, admittedly, large ${\BT}^2$, by blowing up the size of $\CalD$ to even larger proportions.  
It was understood in \cite{Issues, LNS} that the resulting 
two dimensional theory on $\CalD$ is the ${\CalN}=2$ supersymmetric Landau-Ginzburg theory (if instead of $\BT^2$ one takes a Riemann surface $\Sigma$ the original theory was to be partially twisted along $\Sigma$). The target space is the (disconnected) space consisting of the choices of the fluxes ${\bf n}, {\bf m}$ as in \eqref{eq:effcou} and the moduli $(u_{1}, \ldots , u_{r})$ of vacua of the four dimensional theory. The additional degrees of freedom coming from flat abelian connections on ${\BT}^{2}$ are effectively squeezed in the effective metric  and can be neglected for the discussion in this section.

The superpotential (not to be confused with the twisted superpotential) of that theory is
\cite{Issues, LNS}: 
\beq 
{\CalW}_{\bf n, m} = 2\pi {\ii} \sum_{j=1}^{r} \left( n_{j}a^{j} + m^{j}a_{D,j} \right) - \sum_{k=1}^{r} {\beta}_{k} u_{k}
\eeq
where $a^j$ and
\beq
a_{D,j} = \frac{{\partial}{\CalF}}{{\partial}a^j}
\label{eq:dad}
\eeq
are the electric and magnetic special coordinates on the Coulomb branch of the moduli space of vacua.  
Below we recall the more invariant definition of this superpotential in the language of the algebraic integrable systems, which are always behind the special geometry of the Coulomb branch of ${\CalN}=2$ supersymmetric gauge theory in four dimensions \cite{Donagi:1995cf} (see also \cite{Gorsky:1995zq} and \cite{Martinec:1995by}).

Note that the similar flux-induced superpotential, for ${\bar\beta} = 0$ arises in the context of ${\CalN}=1$, $d=4$ theories obtained by engineering using non-compact Calabi-Yau threefold compactifications \cite{TV} and (also for ${\bar\beta}=0$) in the context
of Type II Calabi-Yau fourfold compactifications producing ${\CalN}=2$ two dimensional supergravity theories \cite{GVW}.

\subsection{Algebraic integrable systems}

Let $({\CalX}^{2r}_{\BC}, B^{r}_{\BC}, {\varpi}_{\BC}, {\pi})$ be an algebraic integrable system \cite{Hitchin:1987mz}, with ${\pi}: {\CalX}^{2r}_{\BC} \longrightarrow B^{r}_{\BC}$ defining a Lagrangian fibration  with the fibers $J_{u} = {\pi}^{-1}(u)$, $u \in B^{r}_{\BC}$ being principally polarized abelian varieties.   Let $(u_{1}, \ldots , u_{r})$ denote the basis of global functions on $B^r_{\BC}$, the Hamiltonians of the integrable system. Let ${\beta}_{1}, \ldots , {\beta}_{r}$
denote the corresponding Gibbs chemical potentials (generalised inverse temperatures). 

Let ${\Xi} \subset B^{r}_{\BC}$ denote the discriminant, the set of singular fibers. It is a stratified variety, with the maximal dimension strata of dimension $r-1$. 
Let $u_{*} \in B^{r}_{\BC} \backslash \Xi$  denote the generic point, and 
${\gamma} \in H_{1}(J_{u_{*}}, {\BZ})$ a cycle in the fiber over that point. 
Let $(u, p)$ be a pair consisting of a point $u \in B^{r}_{\BC}$ and a
homotopy class $p$ of a path connecting $u_{*}$ to $u$. We can transport
$\gamma$ over $u$ (the corresponding flat connection on the 
associated vector bundle $\bigcup\limits_{u \in B^{r}_{\BC}} \ H_{1}(J_{u}, {\BC})$ is called
the Gauss-Manin connection), producing a two-dimensional chain $S_{p} \subset
{\CalX}^{2r}_{\BC}$. Define:
\beq
{\CalW}_{\gamma ; u_{*}} (u,p) = {\ii} \int_{S_{p}} {\varpi}_{\BC}  - \sum_{k=1}^{r} {\beta}_{k} u_{k}
\label{eq:suppw}
\eeq
In any integral symplectic basis $A_i$, $B^i$, $i=1,\ldots , r$ in $H_{1}(J_{b_{*}}, {\BZ}) \approx {\BZ}^{2r}$, obeying
\beq
A_{i} \cap A_{j} = B^{i} \cap B^j = 0, \qquad A_{i} \cap B^j = {\delta}_{i}^{j}
\eeq
with the intersection form $\cap$ given by the polarization, we can rewrite \eqref{eq:suppw}
as:
\beq
{\CalW}_{\gamma; u_{*}} (u, p) = 2\pi {\ii}  \sum_{i=1}^{r} \,  \left( n_{i} a^{i}(u) + m^{i} a_{D,i}(u) \right)  - \sum_{k=1}^{r} 
{\beta}_{k} u_{k} \, , 
\label{eq:suppw2}
\eeq
where $a^{i}(u)$, $a_{D,i}(u)$ are the special coordinates on $B^r_{\BC}$, $m^i, n_i \in \BZ$, 
\beq
a^{i}(u) = \frac{1}{2\pi \ii} \oint_{A_i} \, \lambda \, , \qquad a_{D,i}(u) = \frac{1}{2\pi \ii} \oint_{B^i} \, {\lambda}
\eeq
and $\lambda = d^{-1}{\varpi}_{\BC}$. The ambiguity in the definition of $\lambda$ corresponds to the choice of $u_{*}, p$ in \eqref{eq:suppw2}, shifting ${\CalW}_{\gamma; u_{*}}(u,p)$ by a constant.  The fundamental group ${\pi}_{1}( B^{r}_{\BC} \backslash {\Xi} ) $ of the complement $B^{r}_{\BC} \backslash {\Xi}$ acts on $H_{1}(J_{u_{*}}, {\BZ})$ by symplectic transformations via Gauss-Manin monodromy representation.

A celebrated example of such a system is Hitchin's system \cite{Hitchin:1987mz}, 
for which ${\CalX}^{2r}$ is the moduli space of stable Higgs pairs $({\CalP}, {\Phi})$
where ${\CalP}$ is a holomorphic principal $G_{\BC}$ bundle over 
a complex curve $C$, and $\Phi$ is a holomorphic section of the 
bundle $K_{C} \otimes {\mathfrak g}_{\CalP}$. Here $r = (g-1){\rm dim}(G)$, where $g$ is the genus of $C$.

The abovementioned elliptic
Calogero-Moser system is a version of Hitchin system for $G = SL(N)$, 
for a punctured elliptic curve (see \cite{GN2, NH}, and implicitly in \cite{DM}). 

The importance of this example is both the explicit nature of the Hamiltonians and the plethora of simple choices of real slices of the phase space which can be conceivably  
quantized. Then, it was this example (and its degeneration to the periodic Toda chain) which for the first time connected the supersymmetric gauge theory and the algebraic integrable system: namely it is the ${\CalN}=2^{*}$ theory whose low-energy dynamics is governed by the elliptic Calogero-Moser system (this statement is now proven for the $SU(n)$ gauge theories \cite{NO, NP1}). 

The general correspondence is the following \cite{SW4}: the phase space ${\CalX}^{2r}_{\BC}$
is the moduli space of vacua of the ${\CalN}=2$ theory on ${\BS}^{1} \times {\BR}^{1,2}$
with the inverse size of ${\BS}^{1}$ determining the K\"ahler class of the fibers $J_b$. The base $B^{r}_{\BC}$ is the moduli space of vacua of the four dimensional theory. 
The compactification on ${\BS}^1$ and the compactification on ${\BT}^{2} = {\BS}^{1} \times {\BS}^{1}$ are qualitatively similar: in both cases one ends up with the sigma model on 
${\CalX}^{2r}_{\BC}$. 

From the point of view of the classical integrability the special coordinates $a^i$ (or $a_{D,i}$, or some integral linear combinations of those) are the action variables. They are distinguished by the fact that the canonically conjugate angle variables ${\varphi}_{i}$, such that
\beq
{\varpi}_{\BC} = \sum_{i=1}^{r} da^i \wedge d{\varphi}_{i} =   \sum_{i=1}^{r} da_{D,i} \wedge d{\tilde\varphi}^{i}
\eeq
have the rigid periodicity:
\beq
{\varphi}_{i} \sim {\varphi}_{i} + 2\pi \left( n_{i} + {\tau}_{ij} m^{j} \right)\, \qquad
{\tilde\varphi}^{i} \sim {\tilde\varphi}^{i} + 2\pi \left( m^{j} - ({\tau}^{-1})^{ij} n_{i} \right) 
\eeq
with $n_i , m^i \in \BZ$. They are the linear coordinates on the abelian fibers $J_b$, which are isomorphic to ${\BC}^{r}/{\BZ}^{r} \oplus {\tau}{\BZ}^{r}$. If we denote by ${\Gamma}$ the rank $r$ lattice ${\BZ}^r$, in the first description, and
by ${\tilde\Gamma} = {\tau}{\Gamma}$, then the coordinates
${\varphi}$ correspond to the isomorphism ${\BC}^{r} \approx
{\Gamma} \otimes {\BC}$, while ${\tilde\varphi}$ correspond
to ${\BC}^{r} \approx {\tilde\Gamma} \otimes {\BC}$. 

\subsection{Landau-Ginzburg vacua}

Let us now look for the ground states of that effective Landau-Ginzburg theory, i.e. the critical points $u_{c}$ of ${\CalW}_{\gamma ; u_{*}}$. In fact, the derivative $d{\CalW}_{\gamma; u_{*}}$ is $u_{*}$-independent, so the set of critical points is $u_{*}$-independent. Let us assume that $(a^{i})_{i=1}^{r}$ are good local coordinates on $B^r_{\BC}$ near some critical point $u_{c} \in B^r_{\BC}$. Then $d{\CalW}_{\gamma ; u_{*}} = 0$ is equivalent to the system of equations:
\beq
n_{i} + \sum_{j=1}^{r} {\tau}_{ij} m^{j} =  \frac{1}{2\pi\ii} \sum_{k=1}^{r} {\beta}_{k} \frac{{\partial} u_{k}}{{\partial}a^{i}} \, , \qquad i = 1, \ldots , r
\label{eq:qc}
\eeq
What do these equations mean?

\subsubsection{Periodic orbit interpretation}

One obvious interpretation of \eqref{eq:qc} comes from the classical mechanics. The holomorphic function 
\beq
H \, = \, -{\ii} \sum_{k=1}^{r} {\beta}_{k}u_{k} \, : \, {\CalX}^{2r}_{\BC} \, \longrightarrow\, {\BC}
\eeq
generates the Hamiltonian flow, which is linear in the action-angle variables:
\beq
{\dot\varphi}_{i} = 2{\pi} \frac{{\partial}H}{{\partial}a^{i}} \, , \ {\dot a}^{i} = 0 
\label{eq:hamflo}
\eeq
Thus, \eqref{eq:qc} means that the orbit of the flow \eqref{eq:hamflo} closes with the period $1$. 

\subsubsection{Spectral curve interpretation}

An algebro-geometric interpretation of \eqref{eq:qc} is obtained by using the functions $( u_{k} )$ as local coordinates on $B^{r}_{\BC}$:
\beq
d{\CalW}_{\gamma; b_{*}} = 0 \, \Leftrightarrow \, 2\pi{\ii} \sum_{i=1}^{r} \left( n_{i} \frac{{\partial}a^{i}}{{\partial}u_{k}}  + m^{i} \frac{{\partial} a_{D,i}}{{\partial} u_{k}} \right) = {\beta}_{k} \, , \qquad k = 1, \ldots , r
\label{eq:dcw}
\eeq 
Now, suppose the abelian variety $J_b$ is a Jacobian of a (spectral) curve ${\CalC}_{b}$
(as it is the case for the $GL(n)$ Hitchin systems), or its Prym subvariety. 

Then the special coordinates $a^i$, $a_{D,i}$ can be defined also by the periods of a meromorphic $1$-differential ${\tilde\lambda}$ on ${\CalC}_{b}$:
\beq
a^i = \frac{1}{2\pi \ii} \oint_{A_i} {\tilde\lambda} \, , \ a_{D,i} = \frac{1}{2\pi \ii} \oint_{B^i} {\tilde\lambda}
\eeq
for some cycles $A_i , B^i \in H_{1}({\CalC}_{b}, {\BZ})$. The variations ${\varpi}_{k}$of $\tilde\lambda$ 
\beq
{\varpi}_{k} = \frac{\partial}{{\partial} u_{k}} {\tilde\lambda}
\eeq
are holomorphic differentials on ${\CalC}_{b}$. Then \eqref{eq:dcw} states that ${\varpi}_{k}$
has the $C_{\bf m,n} = \sum_{i} n_i A^i + m^i B_i$ period equal to $-2\pi\ii \beta_k$. 

\subsection{$\Omega$-deformed Landau-Ginzburg theory}

We now need to review a few facts about the $\Omega$-deformation. 
It is usually discussed in the context of $A$-type models, starting with the $I-$ and $J$-functions 
in two dimensions \cite{Giv, Ko2}
and the $Z$-function of the four dimensional gauge theory  \cite{N2}. In this paper we'll discuss the less familiar $B$-model case\footnote{I thank M.Dedushenko for bringing to my attention the Ref. \cite{Luo:2014sva} where a similar problem has been considered}, taking \cite{Losev:1992tt, Issues} for inspiration. 
Let us start by recalling the field content and supersymmetry of the topological $B$-model in two dimensions \cite{Vafa:1991uz, Wphases}.

\subsubsection{$B$-model off-shell}

Let $\CalY$ be a complex manifold with the local holomorphic coordinates $Y^{i}$, $i = 1, \ldots , n = {\rm dim}{\CalY}$. The fields of the model are ${\bY}^{i} = Y^{i} + {\psi}^{i} + F^{i}$,  where ${\psi}^{i} \in {\Gamma}\left( {\Omega}^{1}_{\Sigma} \otimes Y^{*}{\CalT}_{\CalY} \right)$, $F^{i}\in {\Gamma}\left( {\Omega}^{2}_{\Sigma} \otimes Y^{*}{\CalT}_{\CalY} \right)$,  and two pairs of boson-fermion scalars valued in $Y^{*}{\overline\CalT}_{\CalY}$, $Y^{\ib}, {\eta}^{\ib}$, and ${\chi}^{\ib}, H^{\ib}$, (with $Y, H$ being bosons, ${\eta}, {\chi}$ being fermions). Geometrically,
$Y = \left( Y^{i} , Y^{\ib} \right) : {\Sigma} \longrightarrow {\CalY}$ is a map of a Riemann surface $\Sigma$ to ${\CalY}$. Define the fermionic nilpotent symmetry (twisted supersymmetry):
\beq
\begin{aligned}
{\delta}Y^{i} = 0\, , \qquad & \ {\delta}{\psi}^{i} = dY^{i} \, , \qquad {\delta}F^{i} = d{\psi}^{i} \\
{\delta}Y^{\ib} = {\eta}^{\ib}\, , \qquad & \ {\delta}{\chi}^{\ib} = H^{\ib} \, ,  \\
{\delta}{\eta}^{\ib} = 0\, , \qquad & \ {\delta} H^{\ib} = 0 \, ,  \\
\label{eq:diffdel}
\end{aligned}
\eeq
where $d$ stands for  the de Rham operator on $\Sigma$. 
The action of the model is defined with the help of a Hermitian metric ${\bf g}_{i\jb}dY^{i} dY^{\jb}$ on $\CalY$, the volume two-form ${\varpi}_{\Sigma}$ on $\Sigma$, and the complex structure on $\Sigma$ (one can trade those for a choice of a metric on $\Sigma$):
\beq
{\bf S} = \int_{\Sigma} {\CalO}^{(2)}_{W} + {\delta}\, \int_{\Sigma}\, \left[ {\bf g}_{i\jb} {\psi}^{i} \wedge \star dY^{\jb}    + {\chi}^{\jb} \left( {\bf g}_{i \jb}  F^{i}   - {\varpi}_{\Sigma} \, {\bar\partial}_{\jb}{\bar W} \right) \right]
\label{eq:bfs}
\eeq
where $W: {\CalY} \to {\BC}$ is a holomorphic function, 
\beq
\begin{aligned}
& W( {\bY}^{i} ) = \sum_{i=0}^{2} \, {\CalO}^{(i)}_{W}\, , \qquad {\CalO}^{(i)}_{W} \in {\Omega}^{i}_{\Sigma} \\
& {\CalO}^{(1)}_{W} = {\psi}^{i} {\partial}_{i}W \\
& {\CalO}^{(2)}_{W} = F^{i} {\partial}_{i}W  + \frac 12  {\psi}^{i} \wedge {\psi}^{j}\, {\partial}^{2}_{ij}W\\
\label{eq:wsup}
\end{aligned}
\eeq
and ${\bar W}$ is a complex conjugate of $W$ in the twisted version of the ${\CalN}=(2,2)$ theory. If we only care about the $\delta$-supersymmetry, i.e. the topological theory, we are free
to deform ${\bar W}$ into anything we like, as long as the action is sufficiently non-degenerate. 

When ${\Sigma}$ has a boundary, the action \eqref{eq:bfs} is not ${\delta}$-invariant:
\beq
{\delta}{\bf S} = \int_{{\partial}{\Sigma}} {\CalO}^{(1)}_{W}
\label{eq:non-inv}
\eeq

\subsubsection{The $\Omega$-deformation of the $B$ model}

Now suppose we have a vector field $V \in Vect({\Sigma})$ which preserves both the complex structure of $\Sigma$ and $\varpi_{\Sigma}$ (i.e. it is an isometry of the corresponding metric). 
The supersymmetry ${\delta}$ can be deformed into the equivariant supersymmetry 
${\delta}_{\hbar}$ which acts as follows:
\beq
\begin{aligned}
{\delta}_{\hbar} Y^{i} = \iota_{V}{\psi}^{i} \, , \qquad & \ {\delta}_{\hbar}{\psi}^{i} = dY^{i} + \iota_{V} F^{i}  \, , \qquad {\delta}_{\hbar} F^{i} = d{\psi}^{i} \\
{\delta}_{\hbar}Y^{\ib} = {\eta}^{\ib}\, , \qquad & \ {\delta}_{\hbar}{\chi}^{\ib} = H^{\ib} \, ,  \\
{\delta}{\eta}^{\ib} = \iota_{V} d Y^{\ib} \, , \qquad & \ {\delta}_{\hbar} H^{\ib} = \iota_{V} d {\chi}^{\ib} \, ,  \\
\label{eq:diffdeleps}
\end{aligned}
\eeq
The action \eqref{eq:bfs} is modified to
\beq
{\bf S}_{\hbar} = \int_{\Sigma} {\CalO}^{(2)}_{W} + {\delta}_{\hbar}\, \int_{\Sigma}\, \left[ {\bf g}_{i\jb} {\psi}^{i} \wedge \star dY^{\jb}    + {\chi}^{\jb} \left( {\bf g}_{i \jb}  F^{i}   - {\varpi}_{\Sigma} \, {\bar\partial}_{\jb}{\bar W} \right) \right]
\label{eq:bfs}
\eeq
where we now assume both ${\varpi}_{\Sigma}$ and $\star$ to be $V$-invariant. 
In the presence of boundary, cf. \eqref{eq:non-inv},
\begin{multline}
{\delta}_{\hbar} {\bf S}_{\hbar} = \int_{\Sigma} d{\psi}^{i} {\partial}_{i} W + F^{i} {\partial}^{2}_{ij}W \, \iota_{V} {\psi}^{j} + {\partial}^{2}_{ij}W \left( dY^{i} + \iota_{V} F^{i} \right) {\psi}^{j}  \\
+ \frac 12 {\partial}^{3}_{ijk} W {\psi}^{i}{\psi}^{j} \iota_{V} {\psi}^{k} = 
\int_{\partial\Sigma} {\psi}^{i} {\partial}_{i}W
\label{eq:dsve}
\end{multline}
Now let us assume ${\partial}{\Sigma}$ to be $V$-invariant and add a term
\beq
\int_{\partial\Sigma} W (Y)\, V^{\vee}
\label{eq:wtr}
\eeq
with one-form $V^{\vee}$ obeying $\iota_{V}V^{\vee} = 1$, $L_{V}V^{\vee} = 0$ on 
$\partial\Sigma$, so that its $\delta_{\hbar}$-variation cancels \eqref{eq:dsve}. The correction \eqref{eq:wtr} is known as Warner term in other contexts. 

Let us now analyze the localization locus for the path integral 
\beq
\int e^{-{\CalS}_{\hbar}}
\eeq
with
\beq
{\CalS}_{\hbar} = {\bf S}_{\hbar} - \int_{\partial\Sigma} W (Y) V^{\vee} 
\label{eq:modac}
\eeq
We want to set the right hand side of \eqref{eq:diffdeleps} to zero:
\beq
dY^{i} = \iota_{V}F^{i}\, , \  d{\psi}^{i} = 0\, , \ \iota_{V}{\psi}^{i} = 
{\eta}^{\ib} = \iota_{V} d{\chi}^{\ib} = 0\, , \ 
H^{\ib} = \iota_{V} dY^{\ib} = 0 \, .
\label{eq:susyloc}
\eeq
The field $H^{i}$
enters $\CalS_{\hbar}$ linearly. Thus we can integrate it out, taking into account the appropriate boundary conditions. This will lead to  the constraint:
\beq
  F^{i}   = {\varpi}_{\Sigma} \ {\bf g}^{i \jb}{\bar\partial}_{\jb}{\bar W}
  \eeq
This constraint implies the following gradient-like equations:
\beq
dY^{i} ={\bf g}^{i \jb}{\bar\partial}_{\jb}{\bar W}
\, \iota_{V} {\varpi}_{\Sigma}
\label{eq:grd}
\eeq
Let us take $\Sigma = {\CalD}$ to be the disk (also known as the cigar), with the radial coordinates $r, \theta$,
$0 \leq r \leq R$, 
with the vector field $V = {\hbar} \partial_{\theta}$, and the ``flat'' volume form:
\beq
{\varpi}_{\Sigma} = r dr \wedge d\theta
\eeq  
The equations \eqref{eq:grd} now take the form:
\beq
\begin{aligned}
\partial_{\theta} Y^{i} = 0 \\
\partial_{r} Y^{i} = {\hbar} r {\bf g}^{i \jb}{\bar\partial}_{\jb}{\bar W} \\
\end{aligned}
\label{eq:yeq}
\eeq
Let us denote by $Y_{R}$ the boundary value of $(Y^{i})$ (it is $\theta$-independent thanks to \eqref{eq:yeq}) and by $Y_{0}$ its value at the center of the disk (the tip of the cigar). 
The boundary term \eqref{eq:wtr} now reads
\beq
\frac{2\pi}{\hbar} W(Y_{R})
\eeq 
whereas the bulk action evaluates to 
\beq
\int_{\Sigma} {\varpi}_{\Sigma} \,  {\bf g}^{i \jb}{\partial}_{i}W {\bar\partial}_{\jb}{\bar W}
= \frac{2\pi}{\hbar} \int_{0}^{R} dr \partial_{r} Y^{i} {\partial}_{i}W = \frac{2\pi}{\hbar} \left( W(Y_{R}) - W(Y_{0}) \right)
\label{eq:bosacloc}
\eeq
so that the path integral reduces to
\beq
\int_{{\Gamma}_{R}} \ e^{- \frac{\CalW}{\hbar}} \ {\tilde\Omega}_{\hbar}
\label{eq:ltc}
\eeq
where ${\CalW} = 2\pi W$, and $\Gamma_R$ is a submanifold in $\CalY$ spanned by the finite-time trajectories of the gradient
vector field, i.e. the solutions
to the equations 
\beq
\partial_{t} Y^{i} = {\hbar} {\bf g}^{i \jb}{\bar\partial}_{\jb}{\bar W}\, , \qquad 0 \leq t \leq R^2
\label{eq:grtrw2}
\eeq
Now let us take the limit $R \to \infty$. Then the only trajectories contributing to the path integral would be those for which the bosonic action \eqref{eq:bosacloc} is finite, which, among other things, implies that $\Vert {\nabla} W \Vert^2 \to 0$ as $R^2 \to \infty$. 
Thus, the boundary value $Y_R$ of the map $Y : \Sigma \to \CalY$ must land at one of the critical points $Y_{*}$ of $W$, i.e. $dW \vert_{Y_{*}}  =0$. 
The center-value $Y_0$, by \eqref{eq:bosacloc} is such, that
\beq
\begin{aligned}
{\rm Im} \left( W(Y_0)/{\hbar} \right) = {\rm Im} \left( W(Y_{R})/{\hbar} \right) \\
{\rm Re} \left( W(Y_0)/{\hbar} \right) \leq  {\rm Re} \left( W(Y_{R})/{\hbar}  \right) 
\end{aligned}
\label{eq:lt1}
\eeq
Thus, $Y_0$ belongs to the Lefschetz timble ${\Gamma}_{Y_{*}}$ of $Y_{*}$, 
i.e. the union of the gradient trajectories, emanating from $Y_{*}$\footnote{I am grateful to E.~Witten 
for discussions about these matters during the work on \cite{NW}. In \cite{WT1, WT2} the related ideas were developed without the use of the $\Omega$-deformation. The advantage of our approach is the possibility of including the Hamiltonians $u_{k}$ into the picture.}. See \cite{Hori:2000ck} for the earlier work on two dimensional ${\CalN}=2$ theories with boundaries. Of course, the equations \eqref{eq:yeq} have appeared in the earlier work \cite{CV92} where they describe the solitons in the Landau-Ginzburg theory. To map our setup to that of Cecotti and Vafa we need to treat the angular coordinate as the time direction, while $r^2/2$ becomes the spatial coordinate. The difference with the \cite{CV92} case is that in our story the spatial interval is only semi-infinite. So only one critical point of $W$ is involved.

The measure ${\tilde\Omega}_{\hbar}$ in \eqref{eq:ltc} comes from the holomorphic
top degree form on $\CalY$. However, for non-compact $\CalY$ it may have a non-trivial $\hbar$-dependence. We shall now get some idea as to what this dependence might be. 

\section{Topological renormalisation group and periods}

Now let us take a break from the gauge theory and try to understand the appearence of Lefschetz thimbles in  
the problem in a more conventional way. 

\subsection{Steepest descent deformations}

The general idea is well-known \cite{AVG}. Let us view the (path) integral
\beq
Z = \int_{\CalF}\ \left[ D{\phi} \right] \ e^{-\frac{S({\phi})}{\hbar}}
\label{eq:initin}
\eeq
of some Euclidean field theory, with $\CalF$ the space of fields obeying the appropriate boundary conditions  as a period integral, i.e. as an integral of the holomorphic top degree form 
\beq
{\Omega}_{\hbar} = \left[ D{\phi} \right] \ e^{-\frac{S({\phi})}{\hbar}}
\eeq
along a middle-dimensional contour $\Gamma \subset {\CalF}^{\BC}$
in the space of complexified fields ${\varphi}$. In \eqref{eq:initin} we have the starting contour $\Gamma_{0} = \CalF \subset {\CalF}^{\BC}$. Now, the value of the integral wouldn't change if we moved the contour while keeping it in the region where the integral converges. In other words, we should not change the large-field, large-momentum asymptotics, but we are free to move the field space contour otherwise. For example, if $V$ is any vector field on ${\CalF}^{\BC}$, $V \in Vect( {\CalF}^{\BC})$, and $g_{t}$ the one-parametric family of diffeomorphisms  of $\CalF^{\BC}$ it generates, then
\beq
\int_{{\Gamma}_{t}} g_{t}^{*}{\Omega}_{\hbar} = Z
\eeq
is $t$-independent. 

\noindent{\bf Proof:} 
\beq
\frac{d}{dt}Z = \int_{{\Gamma}_{t}} g_{t}^{*} Lie_{V}{\Omega}_{\hbar} = 
\int_{{\Gamma}_{t}} g_{t}^{*} d (\iota_{V}{\Omega}_{\hbar} ) = 
\int_{{\Gamma}_{t}} d \left( g_{t}^{*} \iota_{V}{\Omega}_{\hbar} \right) = 0
\eeq
Now, the actual form of the integrand changes with $t$. A popular choice is to take $V$ to be the gradient vector field for ${\rm Re}(S/\hbar)$:
\beq
V = h^{ij} \frac{\partial}{\partial x^{j}}
 \left(  {\rm Re}(S/\hbar) \right) \frac{\partial}{\partial x^{i}}
 \label{eq:gradvf}
\eeq
for some metric $h$ on ${\CalF}^{\BC}$. Outside the set of critical points of $S$ the gradient flow decreases the absolute value of the measure factor $e^{-S/\hbar}$, making their contribution less and less important. As a result, asymptotically, the integral will be dominated by the contribution of the critical points of $S$. Note that the choice of the metric $h$ is arbitrary. Moreover, we can iterate the procedure, by using one vector field $V_{1}$ for some ``time'' $t_{1}$, then another vector field $V_{2}$, e.g. the gradient vector field corresponding to another metric $h_{2}$, for some ``time'' $t_2$, or, more generally, make $V$ ``time''-dependent, e.g. of the form:
\beq
V(t) = h^{ij}(t) \frac{\partial}{\partial x^{j}}
 \left(  {\rm Re}(S/\hbar) \right) \frac{\partial}{\partial x^{i}} \ .
\eeq
Moreover, by taking $h$ to be Hermitian, i.e. of the $(1,1)$ type in the complex structure of ${\CalF}^{\BC}$, we get an additional bonus in that the imaginary part of $S/{\hbar}$
is preserved along the flow:
\beq
{\rm Im} \left( S/{\hbar} \right) = {\rm const} \ .
\label{eq:imp}
\eeq
In this way we map the problem of computing $Z$ given by \eqref{eq:initin} to the problem of computing the partition function of the theory with the $t$-dependent action:
\beq
S_{t} = g_{t}^{*}S \eeq
and the measure given by the restriction of the holomorphic form 
${\Omega}$ onto ${\Gamma}_{t} = g_{t}({\CalF})$. We shall call the $t$-flow on the space of theories the {\it topological renormalisation flow}. 

Let us now discuss to what extent the critical points of $S$ determine the universality classes of the theories. Let us from now on fix a Hermitian metric $h$ (possibly $t$-dependent). Let ${\Gamma}_{*}$
be the union of the trajectories, {\it emanating} from the critical point $x_{*}$, $dS\vert_{x_{*}} = 0$. More generally\footnote{I thank S.~Gukov for a conversation about such situation}, let $F$ be a connected component of the set $Crit(S)$ of critical points of $S$, 
\beq
dS \vert_{f} = 0 \ , \ f \in F
\eeq
We assume $F$ is compact (otherwise our theory might be ill-defined). 
Let ${\gamma}_{\sf f}$, ${\sf f} = 1, \ldots , r_{F}$ 
be a basis of the middle-dimensional homology of $F$. 
Let us choose some representatives ${\hat\gamma}_{\sf f} \subset F$ 
of the corresponding cycles  ${\gamma}_{\sf f}$ . Let ${\Gamma}_{\sf f}$ be the union of all trajectories emanating from the representative ${\hat\gamma}_{\sf f}$. Now, 
define
\beq
Z_{\sf f} = \int_{{\Gamma}_{\sf f}} {\Omega}_{\hbar}\ .
\label{eq:ibapf}
\eeq
The original integral can be expanded:
\beq
Z = \sum_{\sf f} \, n_{\sf f} \, Z_{\sf f} \ ,
\label{eq:zsf}
\eeq
according to the decomposition of the homology class
\beq
\left[ {\Gamma} \right] = \sum_{\sf f} \, n_{\sf f}  \left[ {\Gamma}_{\sf f} \right]
\eeq
in the basis of Lefschetz thimbles. The integers $n_{\sf f}$, in the finite dimensional case, can be computed by the intersection
index
\beq
n_{\sf f} = \# {\Gamma} \cap {\Gamma}_{\sf f}^{\vee}
\eeq
with the dual Lefschetz thimbles ${\Gamma}_{\sf f}^{\vee}$ (defined with the 
help of the gradient flow of $-{\rm Re}(S/{\hbar})$, i.e. the negative of
\eqref{eq:gradvf}). 

In the infinite-dimensional case the intersection theory is subtle. However, one can sometimes argue for vanishing of $n_{\sf f}$ using \eqref{eq:imp}. We should stress
that the topological renormalisation group flow makes the original contour $\Gamma$ asymptotically approach the union of ${\Gamma}_{\sf f}$'s non-uniformly. Namely, $\Gamma$ itself is split into the parts ${\Gamma}_{f}$, which are individually attracted to different Lefschetz thimbles. The levels of the imaginary part ${\rm Im}(S/{\hbar})$ evaluated at the critical points separate these parts $\Gamma_{f}$.

\subsection{Lefschetz thimbles in quantum mechanics}

Let us now apply this formalism to the path integral computing the Euclidean partition function
\beq
{\CalZ} ({\beta}) = {\Tr}_{\CalH} \, 
e^{-\frac{\beta}{\hbar}{\hat H}}
\label{eq:thpf}
\eeq
It can be formally represented as an integral over the space of loops $L{\CalX}^{2r}$ in the classical phase space. The points in $L{\CalX}^{2r}$ are the maps $x: {\BS}^{1} \to {\CalX}^{2r}$, which we shall view as $1$-periodic, $x(s+1) = x(s)$, $x(s) \in {\CalX}^{2r}$
\beq
{\CalZ} ({\beta}) = \int \left[ Dp(s) Dq(s) \right] \, e^{\frac{\ii}{\hbar} \int pdq - \frac{\beta}{\hbar} \int H(x(s)) ds}
\label{eq:piac}
\eeq
where we introduced the Darboux coordinates for the symplectic form $\varpi$ on ${\CalX}^{2r}$: 
\beq
\varpi = \sum_{a=1}^{r} dp_{a} \wedge dq^{a}
\eeq
The details of the definition of \eqref{eq:piac} for non-exact $\varpi$ are discussed in numerous sources so will not be addressed here. It is also well-known 
\cite{Faddeev:1980be} that the domain of integration in \eqref{eq:piac} is not really the loop space of ${\CalX}^{2r}$, rather it is the ill-defined space of loops valued in the space of leaves of some polarization. Of course, as integrals go, the same quantity can be obtained by integration of different measures over different spaces. In fact, this ambiguity leads to the novel symmetries of quantum field theories, Nakajima's algebras
\cite{Na1,Na2,Na3} and their generalizations \cite{N3}. In this paper, however, we are only interested in the enumeration of the possible Lefschetz thimbles, without going into their internal details.  So we shall continue as if we were indeed integrating over $L{\CalX}^{2r}$. 

Let us assume, for simplicity, that the classical phase space $( {\CalX}^{2r} , {\varpi})$ is a real slice
of a complex symplectic manifold $({\CalX}^{2r}_{\BC}, {\varpi}_{\BC})$. Then $L{\CalX}^{2r}$ is a real slice of $L{\CalX}^{2r}_{\BC}$. 

The action $S = {\ii} \oint \sum_{a=1}^{r} p_{a} dq^{a} \ - \ {\beta} \int H(p, q) ds$
is now a holomorphic function on $L{\CalX}^{2r}_{\BC}$. Its critical points are the
$1$-periodic solutions to the Hamilton equations with complex $p$ and $q$, with the Hamiltonian 
$-{\ii}{\beta}H(p,q)$:
\beq
{\dot q} = - {\ii}{\beta} {\partial}_{p} H\, , \ {\dot p} = {\ii}{\beta} {\partial}_{q} H
\label{eq:pqhb}
\eeq
Now the circle closes: assume $\left( {\CalX}^{2r}_{\BC}, {\varpi}_{\BC} \right)$ is an algebraic integrable system. One can generalize the problem \eqref{eq:piac} to include several Hamiltonians ${\beta}H \to \sum_{k=1}^{r} {\beta}_{k}u_{k}$. The $1$-periodic complex
loops solving \eqref{eq:pqhb} are exactly the solutions to \eqref{eq:qc}. 

Thus, the Lefschetz thimbles of quantum mechanical partition function are in one-to-one correspondence with the critical points of the superpotential ${\CalW}_{\bf m, n}$!
Well...almost. The critical point of ${\CalW}_{\bf m,n}$ defines not just one periodic
solution of \eqref{eq:pqhb}, but the whole torus $J_{b}$  ($b \in B^r$
is specified by fixing the values of all Hamiltonians $u_k$) of those solutions: 
the $s=0$ value of the angle variables $({\varphi}_{i})_{i=1}^{r}$ can be chosen 
arbitrarily.  Of these $r$ complex moduli one real modulus is simply the translation
in $s$. 

\subsection{Twists}

Suppose $({\CalX}^{2r} , {\varpi}, H)$ has a symmetry group $G$, i.e. 
is invariant under the action of some group $G$. 
Suppose the symmetry extends to the quantization, 
as the unitary representation of $G$
in ${\CalH}$, commuting with ${\hat H}$. The partition function \eqref{eq:thpf}
generalizes to the character 
\beq
{\CalZ}_{g} ({\beta}) = {\Tr}_{\CalH} \, \left( g e^{-\frac{\beta}{\hbar}{\hat H}} \right)
\label{eq:czgbh}
\eeq
The character-valued partition function \eqref{eq:czgbh} has the path integral representation, where one integrates (again, in the \cite{Faddeev:1980be} sense) 
over the space of twisted loops:
\beq
x(s+1) = g\cdot x(s) \ .
\label{eq:twlps}
\eeq
 The integral only depends on the conjugacy class of $g$ in $G$ (the transformation
 $g \mapsto h^{-1}gh$ can be undone by $x(s) \mapsto h^{-1}\cdot x(s)$). 
What are the Lefschetz thimbles for this problem?

Let us assume, again, that $({\CalX}^{2r}_{\BC}, {\varpi}_{\BC})$ is an algebraic integrable
system, and that the action of $G$ on ${\CalX}^{2r}$ extends to $({\CalX}^{2r}_{\BC}, {\varpi}_{\BC})$, where it acts by holomorphic symplectic transformations
preserving the integrals. 

Suppose $G$ is finite. The Hamiltonians $u_k$, the action variables $a^i$ and their duals $a_{D,i}$
are $G$-invariant. The $G$-action on the angle variables is highly constrained. 
Let us assume that $G$ acts by shifts: 
\beq
g\, : \, {\varphi}_{i} \mapsto {\varphi}_{i} + \frac{2\pi}{N(g)} \left( n_{i}(g) +  \sum_{j=1}^{r} {\tau}_{ij} m^{j}(g) \right) 
\label{eq:dscrac}
\eeq
where $N(g)$ is the order of $g \in G$ (recall that by Cayley's theorem $G \subset S(|G|)$
so every conjugacy class belongs to a cyclic subgroup ${\BZ}/N(g){\BZ}$ of sufficiently large order), $n_{i}(g), m^{j}(g)$ are integers defined modulo
$N(g) {\BZ}$. Then \eqref{eq:twlps} condition translates
to the following modification of \eqref{eq:qc}:
\beq
n_{i} + \frac{n_{i}(g)}{N(g)} +  \sum_{j=1}^{r} {\tau}_{ij} \left( m^{j} + 
\frac{m^{j}(g)}{N(g)} \right) = \frac{1}{2\pi\ii} \sum_{k=1}^{r} {\beta}_{k} \frac{{\partial} u_{k}}{{\partial}a^{i}} \, , \qquad i = 1, \ldots , r
\label{eq:qc2}
\eeq
Of course, the solutions  to \eqref{eq:qc2} define closed loops on the quotient ${\CalX}^{2r}_{\BC}/G$.  

\subsection{The rest of the computation}

To compare with \cite{Col,Pol1} we need to classify the critical points
$u_{c} = u_{i, \bf m, n}$ of ${\CalW}_{\bf m,n}$, compute the multiplicities $n_{i, \bf m , n}$, evaluate the critical value ${\CalW}_{\bf m,n} (u_{i, \bf m, n})$ of the action, estimate the one-loop corrections, and 
perform the sum over ${i, \bf m, n}$. We leave this for future work. 

Let us make a couple of comments. 

\begin{enumerate}

\item{}
For the solution of 
\eqref{eq:qc} the values of $\bf m, n$ are basis-dependent. However, their greater common 
divisor (g.c.d.) $N$ is basis-independent. Its invariant meaning is the multiplicity of the $1$-cycle $[C_{i, \bf m,n}] \in H_{1}(J_{u_{c}}, {\BZ})$ on the abelian variety, which is represented by the periodic orbit $C_{i, \bf m,n}: S^{1} \to J_{u_{c}}$ of the Hamiltonian vector field. In other words, $[C_{i, \bf m,n}] = N [C_{0}]$, where $[C_{0}] \in H_{1}(J_{u_{c}}, {\BZ})$ is a primitive class.

Another invariant integral data, apart from the g.c.d. $N$, comes from the set of vanishing cycles. Let us assume ${\beta}_{k}$ are so large, that $u_{c}$ is close to the discriminant locus $\Xi_{\alpha} \subset \Xi$, where the cycle $[C_{\alpha}]$ vanishes. For the details of the construction of the basis of vanishing cycles see \cite{AVG}.  The intersection number (it is defined using the polarization of the abelian variety) $n_{\alpha} = \# [C] \cap [C_{\alpha}]$ is well-defined. 
This is what we might call the number of instanton-antiinstanton pairs of type $\alpha$. 
The `dual' cycle $[C_{\alpha}]^{\vee}$ is defined up to the addition of $[C_{\alpha}]$ (Picard-Lefschetz theory) \cite{AVG}, so that the dual numbers $m_{\alpha}$ are defined modulo $n_{\alpha}$. 

\item{}

The critical locus of $S$ containing the orbit $C_{i, \bf m,n}$ is the abelian variety $J_{u_{c}}$. The middle-dimensional homology $H_{r}(J_{u_{i, \bf m,n}}, {\BZ})$ group has rank 
$(2r)!/r!^2$. It is a representation of the monodromy group, which is a subgroup of $Sp(2r, {\BZ})$. The choice of the cycles ${\gamma}_{i, \bf m,n} \in H_{r}(J_{u_{i, \bf m,n}}, {\BZ})$ is constrained by the monodromy equivariance property. In the $r=1$ case the choice is simple: the $1$-cycle is represented by the orbit itself.

\item{}

In the instanton gas prescription the enthropy prefactor ${\beta}^{N}/N!$ of the configuration of $N$ instantons and anti-instantons, which comes from the integration over the ``zero modes'', is important in converting the series into the exponential of the energy splitting. However, as we argued above, the number of zero modes of a true critical point of the analytic continued action is $N$-independent. Thus, we must get the ${\beta}^{N}/N!$ from the one-loop fluctuation determinant. Somehow the $N$ zero modes of the approximate
solution should flow to $N$ low-lying eigenvalues of the linearization of the equations of motion around the true solution. Since the linearized equations are the second order differential equation on a circle of periodicity $\beta$ while the (matrix) coefficients of the 
equation have the periodicity ${\beta}/N$, one can construct the eigenvectors
by taking the Bloch solutions with the Bloch phases given by the $N$'th roots of unity. 

\item{}
 
The one-loop correction is computed by the regularized determinant of 
the second quadratic form of the action $S$ expanded about the critical point. 
As we have complexified the space of fields, we should view it as an infinite-dimensional quadratic form. The convergent Gaussian integral along the Lefschetz thimble ${\Gamma}_{i, \bf m,n}$ has a phase, 
which comes from the restriction of the holomorphic top form $\Omega_{\hbar}$  onto a
real subspace of the tangent space to $L{\CalX}^{2r}_{\BC}$ at $C$, spanned by the 
tangent vectors to the representative ${\hat\gamma}_{i, \bf m,n}$, as well as the eigenvectors of ${\delta}^{2} {\rm Re}(S/{\hbar})$ with positive eigenvalues. 
Of course, the multiplication by $\ii$ maps such an eigenvector to the eigenvector with the opposite eigenvalue. 

We should however remember that the integration domain is not precisely $L{\CalX}^{2r}$. 
If we approximate the loop by an $K$-gon, with a very large $K$, the action
$S$ is approximated by \cite{Faddeev:1980be}:
\beq
S = - {\ii} \sum_{k=1}^{K} p_{a,k} (q^{a}_{k+1}-q^{a}_{k}) + \frac{\beta}{K} \sum_{k=1}^{K} 
H (p_{k},q_{k}) \, , \
\label{eq:slat}
\eeq
with $q_{K+1}= q_{1}$. 
This is to be compared, e.g. to the integral of $-{\ii}{\varpi}$ over the polygon in the ${\BR}^{2r}$
with the vertices $(p_{k}, q_{k})$. In computing the ${\delta}^{2}S$ one should keep in mind  the choice of the polarization.

\end{enumerate}

\section{Examples of the models}

In this section we consider a few specific examples. We start with the systems with one degree of freedom: Heun system (Gaudin on $4$-points), Lam{\'e} system, and the anharmonic oscillator. 

\subsection{One degree of freedom}

Let us start with the systems describing a one dimensional particle (or a spin degree of freedom): let us assume the Hamiltonian is quadratic in momentum:
\beq
H (p,x) = \frac{1}{2} A(x) p^2 + B(x) p + C(x) 
\label{eq:hpxabc}
\eeq
with some functions $A,B,C$. Such system is an algebraic integrable system
if the energy level set $H(p,q) = E$ is an elliptic curve (perhaps missing
a few points). For example, one can choose $A$ to be degree $4$
polynomial \cite{Turbiner:2016aum}. 

By going to the coordinate ${\tilde x} \sim \int dx/\sqrt{A(x)}$ and performing
a similarity transformation of the wavefunction the Hamiltonian 
\eqref{eq:hpxabc} can be mapped to the standard non-relativistic form
\beq
{\tilde H}({\tilde p},{\tilde x}) = \frac{1}{2} {\tilde p}^{2} + U({\tilde x}) \ .
\label{eq:nonrel}
\eeq
for some function $U({\tilde x})$. 
\subsubsection{Gaudin model}

Let us start with a $n=4$-point $SL(2)$ Gaudin 
model: the phase space ${\CalX}^{2}_{\BC}$ is the complex symplectic
quotient:
\beq
{\CalX}^{2}_{\BC} = \left( {\CalO}_{{\nu}_{1}} \times \ldots \times {\CalO}_{{\nu}_{n}}
\right) // SL(2, {\BC})
\eeq
where
\beq
{\CalO}_{\nu} = \left\{ \, ( x, y, z ) \  | \   x^2 +  y^2 +  z^2 = {\nu}^{2} \,  \right\}
\eeq
with the symplectic form 
\beq
{\varpi}_{\nu} = \frac{dx \wedge dy}{2 z}
\eeq
The space ${\CalX}^{2}_{\BC}$ can be identified with the space
of meromorphic ${\mathfrak{sl}}(2, {\BC})$-valued $1$-forms
\beq
{\phi}(w) = \sum_{a=1}^{n} {\phi}_{a} \frac{dw}{w-w_{a}} 
\label{eq:mhf}
\eeq
where
\beq
{\phi}_{a} = \left( \, \begin{matrix} x_{a} & y_{a} + {\ii} z_{a} \\
y_{a} - {\ii} z_{a} & - x_{a} \end{matrix} 
\, \right)
\label{eq:resa}
\eeq 
obey 
\beq
\sum_{a= 1}^{n} {\phi}_{a} = 0 \, , \ 
\eeq
and we identify $({\phi}_{1}, \ldots , {\phi}_{n}) \sim 
(g^{-1}{\phi}_{1}g, \ldots , g^{-1}{\phi}_{n}g)$, for $g \in SL(2, {\BC})$. 
The positions $w_{1}, w_{2}, w_{3}, w_{4}$ of the poles are the parameters
of the model. Actually, only the cross-ratio 
\beq
{\qe}  = \frac{w_{2}-w_{1}}{w_{3}-w_{1}} \frac{w_{3}-w_{4}}{w_{2}-w_{4}}
\eeq
is relevant (for $n>4$ there will be $n-3$ parameters). However, sometimes it is convenient to use the redundant
parameterization, i.e. in exploring various degenerations, including
the one to the anharmonic oscillator. For now, however, we'll choose $w_{1} = 0, \, w_{2} = {\qe}, \, w_{3} = 1, \, w_{4}={\infty}$. Define the spectral curve ${\CalC}$:
\beq
{\Det} \left( {\rho} - {\phi}(w) \right)  = 0
\eeq
which sits in the cotangent bundle $T^{*}{\BC\BP}^{1}$ to ${\BC\BP}^{1} \ni z$ with the fibers over $0, {\qe}, 1, \infty$ deleted. Explicitly:
\begin{multline}
{\rho}^{2} = \sum_{a=1}^{4} \frac{{\nu}_{a}^{2}}{(w-w_{a})^{2}} + \frac{H_{a}}{w-w_{a}}
= \\
\frac{{\nu}_{1}^{2}}{w^{2}} + \frac{{\nu}_{2}^{2}}{(w-{\qe})^{2}} + \frac{{\nu}_{3}^{2}}{(w-1)^{2}}
+ \frac{{\nu}_{4}^{2} - {\nu}_{1}^{2} - {\nu}_{2}^{2} - {\nu}_{3}^{2}}{(w-1)(w-{\qe})} 
+ \frac{H {\qe}({\qe}-1)}{w(w-1)(w-{\qe})} \, 
\label{eq:Hg4}
\end{multline}
where  $H$ is the Hamiltonian of our system. There is a nice geometric picture of the phase space ${\CalX}^{2}$ as the complexification of the space of closed polygons (quadrangles) 
with fixed (complex) lengths of the edges (${\nu}_{a}$'s) in the three dimensional Euclidean space, viewed up to the isometries. The advantage of the complex situation is that there is no need
for the triangle inequalities: all complex lengths are good. Let us label the vertices
by $p_{a} \in {\BC}^{3}$, $a = 1, \ldots , 4$, so that 
\beq
p_{a+1} - p_{a} = (x_{a}, y_{a}, z_{a}) \, , \quad a = 1, \ldots , 4 \, , \quad p_{a+4} = p_{a}
\eeq
The Darboux coordinates
on this space can be chosen to be the length $p = {\ell}_{12}$ of the diagonal connecting $p_{1}$ and $p_{3}$, and the dihedral angle $x = {\vartheta}_{12}$ between the triangles $p_{1}p_{2}p_{3}$
and $p_{3}p_{4}p_{1}$. 
The Hamiltonian $H$ \eqref{eq:Hg4} is given by:
\beq
H = \frac{{\Tr}({\phi}_{1}{\phi}_{2})}{{\qe}} + \frac{{\Tr}({\phi}_{2}{\phi}_{3})}{{\qe}-1} \ .
\label{eq:Hg4i}
\eeq
Using
\beq
{\ell}_{ab}^{2} = (x_{a} + x_{b})^2 + (y_{a}+y_{b})^{2} + (z_{a} + z_{b})^{2}\, ,
\eeq
and  
\beq
{\rm cos}({\vartheta}_{12}) = \frac{{\Tr} [{\phi}_{1}, {\phi}_{2}][{\phi}_{3},{\phi}_{4}]}{\sqrt{\left( {\Tr} [{\phi}_{1}, {\phi}_{2}]^2 \right) \left( {\Tr} [{\phi}_{3}, {\phi}_{4}]^2 \right)}}
\eeq
we can rewrite \eqref{eq:Hg4i} as:
\begin{multline}
{\qe} (1-{\qe}) H = (1 - {\qe}) \left( p^{2} - {\nu}_{1}^{2} - {\nu}_{2}^{2} \right) + 
\frac{\qe}{2p^2}  \left( p^2 - {\nu}_{1}^{2} + {\nu}_{2}^{2} \right)  \left( p^2 + {\nu}_{3}^{2} - {\nu}_{4}^{2} \right)  + \\
+ \frac{{\qe}\, {\rm cos}(x)}{2p^2} \sqrt{ \left(  \left( p^{2} - {\nu}_{1}^{2} - {\nu}_{2}^{2} \right)^{2} - 4 {\nu}_{1}^{2}{\nu}_{2}^{2} \right) \left( \left( p^{2} - {\nu}_{3}^{2} - {\nu}_{4}^{2} \right)^{2} - 4 {\nu}_{3}^{2}{\nu}_{4}^{2} \right)  } 
\label{eq:Hg4ii}
\end{multline}
See \cite{Turbiner:2016aum} for other realizations of this model. 

The simplest presentation of the Hamiltonian \eqref{eq:Hg4i} is in terms of the separated variables. The specification of that approach to the system with one degree of freedom will produce, upon the usual quantization, the Hamiltonian 
\beq
{\hat H} = - {\hbar}^2 \, x(x-1)(x-{\qe}) \frac{\partial^2}{{\partial x}^2} +  
\frac{{\Delta}_{1} {\qe}}{x} + \frac{{\Delta}_{2} {\qe}({\qe}-1)}{x-{\qe}} + 
\frac{{\Delta}_{3}(1-{\qe})}{x-1}
+ {\Delta}_{4}  x 
\label{eq:Heun}
\eeq
which can be mapped to the standard non-relativistic form \eqref{eq:nonrel} by choosing
$\tilde x$ to live on the elliptic curve:
\beq
y^2 = 4x(x-1)(x-{\qe}), \qquad d{\tilde x} = dx/y
\eeq
The model \eqref{eq:Heun}
is important since it is Bethe/gauge dual to the four dimensional $SU(2)$ gauge theory with $N_f = 4$ fundamental hypermultiplets. 
The parameters ${\Delta}_{i} = {\nu}_{i}({\nu}_{i}-\hbar)$ are related to the masses of four quarks. The parameter $\qe$ is determined by the microscopic gauge coupling.

\subsection{Anharmonic oscillator}

This is the model we shall discuss in more detail: a 
particle in the one-dimensional potential
\beq
U(q) = \frac{\lambda}{4}(q^2 - v^{2})^{2}
\label{eq:u1ofx}
\eeq
It can be viewed as an irregular limit of the general Gaudin model. 
It is not Bethe/gauge dual to any four dimensional gauge theory with Lagrangian description. However 
it can be obtained by a mass deformation of one of the 
non-trivial rank $1$ ${\CalN}=2$ superconformal fixed points \cite{Argyres:2010py}. 

The complexified (and compactified) energy level set
\beq
H(p,q) = \frac{1}{2} p^{2} + U(q) = E
\label{eq:encurv1}
\eeq
is an elliptic curve. Let $U_{0} = U(0) = {\lambda}v^{4}/4$. Let us choose the parametrization:
\beq
p = {\ii} \sqrt{2U_{0}} \, {\rho}\, , \qquad q = v \, {\xi}\, , \qquad E = U_{0}\, {\ep} 
\label{eq:param}
\eeq
so that the curve \eqref{eq:encurv1} becomes: 
\beq
{\rho}^{2} = ({\xi}^{2}-1)^{2}-{\ep}
\label{eq:encurv2}
\eeq
The complex phase space ${\CalX}^{2}_{\BC}$ is a partial compactification of the naive space ${\BC}^{2}$
of complex coordinates $q$ and momenta $p$, where we add the locus (two copies of $\BC$) where $q$ and $p$
go to infinity, as:
\beq
p \sim \pm {\ii}\sqrt{\frac{\lambda}{2}} \left( q^2 - v^2  -  \frac{2E}{{\lambda} q^2} + \ldots \right)
\eeq
This compactification is physically natural, since it takes only a finite time to reach infinity
while moving in the inverted potential. Let us denote the compactified curve by ${\CalE}_{\ep}$. The symmetry $G = {\BZ}_{2}$ acting by 
\beq
(p,q) \mapsto (-p,-q)
\label{eq:invol}
\eeq preserves both the symplectic form and the Hamiltonian $H(p,x)$. 

The elliptic curve ${\CalE}_{\ep}$ \eqref{eq:encurv2} maps $2:1$ to another elliptic curve $E_{\ep}$, which sits in the quotient ${\CalX}^{2}_{\BC}/{\BZ}_{2}$:
\beq
y^2 = 4 x( (x-1)^2 - {\ep} ) 
\label{eq:encurv3}
\eeq
where
\beq
y = 2{\rho}{\xi}\, , \qquad x = {\xi}^{2} \, .
\label{eq:isogeny}
\eeq
Such map is called an isogeny. As long as ${\ep} \neq 1$ the fixed points of the involution \eqref{eq:invol} do not belong to \eqref{eq:encurv2}. Thus, \eqref{eq:invol} acts by a half-period shift, as can be seen explicitly below. 

We thus have represented both ${\CalX}^2_{\BC}$ and ${\CalX}^2_{\BC}/{\BZ}_{2}$
as the algebraic integrable systems, with the same base $B^{1}_{\BC} \approx {\BC}$, with the coordinate $\ep$. The fiber $J_{\ep}$ is the curve \eqref{eq:encurv2}  ${\CalE}_{\ep} \subset {\CalX}^{2}_{\BC}$ and the curve $E_{\ep} \subset {\CalX}^{2}_{\BC}/{\BZ}_{2}$, respectively. 

We can now analyze our problem of finding the periodic and ${\BZ}_{2}$-twisted periodic orbits on ${\CalX}^{2}_{\BC}$. First, let us choose a basis in $H_{1}({\CalE}_{\ep}, {\BZ})$.  
 Suppose $ | \ep | \ll 1$. Let 
 \beq
 {\xi}_{\pm} = \left( 1\pm \sqrt{\ep} \right)^{\frac 12}  \approx 1 \pm \frac 12 \sqrt{\ep}
 \eeq
 We choose the $A$-cycle to be the cycle which circles around the cut ${\left[ {\xi}_{-} , {\xi}_{+}  \right]}$ on the $\xi$-plane, while the $B$-cycle
 is the double cover of the interval connecting the two cuts: ${\left[ - {\xi}_{-}, {\xi}_{-} \right]}$. The $A$-cycle vanishes when ${\ep} \to 0$. The monodromy under ${\ep} \mapsto e^{+2\pi \ii} {\ep}$ takes ${\xi}_{-}$ to ${\xi}_{+}$, the cycle $A$ to itself, and 
 \beq
 B \mapsto B + 2A
 \label{eq:bacycles}
 \eeq
 since the end-points of both small cuts ${\left[ \pm{\xi}_{-}, \pm{\xi}_{+} \right]}$
 get exchanged, so that ${\left[-{\xi}_{-}, {\xi}_{-} \right]}$ turns into 
 \beq
 \left[- {\xi}_{+}, {\xi}_{+} \right] \  = \
  \left[- {\xi}_{+}, - {\xi}_{-} \right] \ \cup \  \left[ - {\xi}_{-}, {\xi}_{-} \right]  \ \cup \ \left[ 
 {\xi}_{-}, {\xi}_{+} \right] \ .  \eeq
 The corresponding periods $\frac{1}{2\pi} \oint pdx $ are given by (with ${\tilde\ep} = {\ep}/64$):
 \beq
 \begin{aligned}
&  a = \frac{{\ii}v\sqrt{2U_{0}}}{{\pi}} \int_{{\xi}_{-}}^{{\xi}_{+}} \ {\rho}d{\xi} \ = \\
& \qquad\qquad = \ -16v\sqrt{2U_{0}} {\tilde\ep}  \left( 1 + 6 {\tilde\ep}  +  140 {\tilde\ep} ^2 + 4620 {\tilde\ep} ^3 \ldots \right)  \, , \\
& \\
& a_{D} =  \frac{{\ii}v\sqrt{2U_{0}}}{{\pi}} \int_{-{\xi}_{-}}^{{\xi}_{-}}\ {\rho}d{\xi}  = \\
& \qquad\qquad  = \  \frac{2{\ii}S_{0}}{2\pi} + \frac{2a}{2{\pi}{\ii}}
\left(  {\rm log}\left( {\tilde\ep} \right) - 1  + 23  
{\tilde\ep}  +  612 {\tilde\ep} ^2 + \ldots \right) \\
& \qquad\qquad \\
\label{eq:aadper}
\end{aligned}
\eeq 
where
\beq
2S_{0} = \frac{8v \sqrt{2U_{0}}}{3} 
\label{eq:twoinst}
\eeq 
is twice the instanton action. 
Note the $2$ in the numerator of $\frac{2a}{2\pi \ii}$ in \eqref{eq:aadper} is the same $2$ as in the monodromy transformation \eqref{eq:bacycles}, and the same $2$ as in \eqref{eq:twoinst}. 
The period of motion along the orbit represented by the cycle $mA + n B$ with $m,n \in {\BZ}$, $m \sim m +  2n$ is given by:
\begin{multline}
{2\pi\ii} \frac{\partial}{\partial E} \left( m a  +  n a_{D} \right) = \\
\frac{{\ii}{\pi}m + n \left( {\rm log}\left( {\tilde\ep} \right) + 40 {\tilde\ep} + 1076 {\tilde\ep}^{2} + \ldots \right) }{\omega}  \cdot \left( 1 + 12 {\tilde\ep}  + 420 {\tilde\ep} ^{2} + \ldots \right) \label{eq:mnper}
\end{multline}
Where
\beq
  {\omega} = \sqrt{U''(v)} = \sqrt{2{\lambda}v^{2}} = 2 \frac{\sqrt{2U_{0}}}{v} 
\eeq
is the frequency of small oscillations near the minimum of the potential \eqref{eq:u1ofx}. In  quantum field theory language $\omega$ is the mass of the perturbative quantum of the field $x(t)$. 
We see that for real $\ep$ the contribution of the $B$ cycle to the period is real (it corresponds to the classically allowed motion in the inverted potential) while the contribution of the $A$ cycle is imaginary. But $\ep$ does not have to be real:
in fact, by setting \eqref{eq:mnper} to be equal to $\beta$ we get, for $\beta \to +\infty$:
\beq
{\tilde\ep} \, =  \, e^{\frac{2{\pi\ii}m}{N}}\, {\exp} \left( - \frac{2{\beta}{\omega}}{N} \right)  + \ldots
\label{eq:epmn}
\eeq
for $n = - |n| < 0$, $N = 2|n|$. The neglected terms are exponentially suppressed. 
The critical value of the action  on such a solution is
equal to 
\beq
{\CalW}_{m,n} = 2{\pi\ii}\left( 1 - E \frac{\partial}{\partial E} \right) \left( m a + n a_{D} \right) =  \ N S_{0} \cdot \left( 1  - 12 \, e^{\frac{2{\pi\ii}m}{N}}\, {\exp} \left( - \frac{2{\beta}{\omega}}{N} \right) + \ldots \right)
\label{eq:cwmn}
\eeq
Now let us discuss the modification of this result for the ${\BZ}_{2}$ twisted trajectories. 
They project to the closed loops on $E_{\ep}$ which are not in the image of closed loops
on ${\CalE}_{\ep}$ under the isogeny \eqref{eq:isogeny}. The first homology 
group of $E_{\ep}$ is generated by the ${\CalA}$-cycle, which circles around the cut
$\left[ {\xi}_{-}^{2}, {\xi}_{+}^{2} \right]$ on the $x$-plane, and the ${\CalB}$-cycle, which is the double cover of the interval $\left[ 0, {\xi}_{-}^{2} \right]$. The isogeny \eqref{eq:isogeny} maps $A$ to ${\CalA}$ and $B$ to $2{\CalB}$. Accordingly the monodromy ${\ep} \mapsto e^{2\pi \ii}{\ep}$ maps ${\CalB} \mapsto {\CalB} + {\CalA}$. 
An analogous analysis will give the result \eqref{eq:epmn}, \eqref{eq:cwmn} with odd $N$. 

Which pairs $(m,N)$ with $m = 0, \ldots , N-1$ actually contribute to the path integral?
In the phase space formulation \eqref{eq:piac} of the path integral the exponential has, for real $\hbar$, an imaginary part $\propto \frac{1}{\hbar} \oint pdq$. The real contour $L{\CalX}^{2r}$ is stratified by the value of $I = \frac{1}{\hbar} \oint pdq$, so that the flow lines of the antigradient
$- {\nabla}{\rm Re}(S/{\hbar})$ emanating from the critical point $(m,N)$ may cross it at the stratum with $I \approx  -12 N S_{0} e^{-\frac{2\beta\omega}{N}} {\rm sin} \left( 2{\pi}m/N \right)$. So, at first sight all the critical points should contribute. However, if we integrate out $p$ and represent the thermal partition function as
the integral over real periodic trajectories $q(t) = q(t+{\beta})$ with the real Euclidean action
\[ \int_{0}^{\beta} \, dt \, \left( \frac 12 {\dot q}^{2} + U(q) \right) \, ,  \]
The gradient flow deforming the contour of integration over real loops will not change the imaginary part of the action \eqref{eq:imp}. Then, as only the
pairs $(0,n)$ and $(n,2n)$ give real critical values (at least in the approximation we are working), so it appears only the ${\Gamma}_{0,n}$ and ${\Gamma}_{n,2n}$ Lefschetz thimbles will contribute to the thermal partition function, at real $\beta$. 

At any rate, once  we take the parameters of the problem (slightly) complex, e.g. in comparing our formalism with \cite{BW},  then all $(m,n)$ pairs play a role.

\subsection{Generalization to several degrees of freedom}

\subsubsection{$g$ identical particles in a polynomial potential}

In our previous discussion the fact that the potential $U(q)$ is quartic seems to play a crucial role. If $U(q)$ is degree $2g+1$ or degree $2g+2$ polynomial, with $g > 1$, then the curve $H(p,q) = E$ is hyperelliptic of genus $g$. The classical motion in the complexified phase space is still a linear flow
in the ``coordinate''
\beq
\int \frac{dq}{p}
\eeq
which has $2g$ periods (the differential $dq/p$ is holomorphic). Fixing the period of motion to be equal to $\beta$ is not enough to fix all the moduli of the curve $H(p,q) = E$, but
should be enough to fix $E$. 
The motion of one particle in such a potential may still be amenable to the analysis like we did for the anharmonic oscillator, but it is not an algebraic integrable system, and is not likely to correspond to some gauge theory. However, the motion of $g$ identical particles is \cite{Sklyanin:1991ss, GNR}. The corresponding abelian variety is the Jacobian of the curve. The integrals of motion are the $g$ parameters $u_{1}, \ldots , u_{g}$ of the curve, which must be chosen in such a way that the derivatives of the Liouville one-form $pdq$ with respect to $u_{k}$
are holomorphic differentials. This is precisely the setup of Seiberg-Witten theory \cite{SW1, SW2}. 

\subsubsection{$N$-particle Toda chain}

The system of $N$-particles governed by the Hamiltonian
\beq
{\hat H} = - \frac{{\hbar}^2}{2} \sum_{i=1}^{N} \frac{\partial^2}{{\partial} q_{i}^{2}} + {\Lambda}^{2} \sum_{i=1}^{N} e^{q_{i}-q_{i+1}}
\label{eq:pertod}
\eeq
where $q_{N+1} = q_{1}$, is famously dual to the pure ${\CalN}=2$ super-Yang-Mills with gauge group $SU(N)$. It was observed in the classical limit in \cite{Gorsky:1995zq, Martinec:1995by}, in \cite{Braverman:2004vv, Braverman:2004cr} in the formal quantization, in \cite{NS3} in the actual $L^2$-quantization on the real line $q_{i} \in {\BR}$. It is also possible to study the eigenvalue problem \eqref{eq:pertod} where $q_{i} \in {\ii}{\BR}/2\pi \BZ$, however new phenomena arise in this case, notably the non-perturbative splitting of the levels for which our formalism is being developed (see \cite{Jeong} for the current status of the problem for $N=2,3$, and \cite{Gorsky:2017ndg} for new developments). 
  
We expect to see our non-linear superpositions of instantons and antiinstantons in the quantum mechanical model in the limit where all Gibbs potentials $\beta_k \to \infty$. 
The corresponding spectral curve
\beq
Y + \frac{{\Lambda}^{2N}}{Y} = P_{N}(x) = x^{N} + u_{1}x^{N-1} + \ldots + u_{N}
\label{eq:spctoda}
\eeq 
becomes maximally degenerate. These are the equilibrium points of Toda chain. The corresponding values of $u_k$'s correspond to the vacua of ${\CalN}=1$ theory which is obtained from ${\CalN}=2$ by soft superymmetry breaking via a superpotential deformation.    
At such a point $n-1$ cycle vanishes. So we expect the critical points to be enumerated by the collections of integers
${\bf m, n}$ with ${\bf n} = (n_{1}, \ldots, n_{N-1})$ and ${\bf m} = (m_{1}, \ldots, m_{N-1})
$ with $m_i$'s defined modulo some lattice generated by $n_i$'s. 

\subsubsection{$n$-point Gaudin system}

The maximal degeneration, which should occur when all ${\beta}_{k} \to \infty$ for $k = 1, \ldots, n-3$:
\beq
{\rho}^2 = \frac{B^{2} (w) ({\alpha}w^2 + {\beta}w + {\gamma})}{R(w)^2}
\eeq
where
\beq
R(z) = \prod_{i=1}^{n} (w  - w_{i})
\eeq
with degree $n-3$ monic polynomial $B(w)$. The $n$ parameters ${\alpha}, {\beta}, {\gamma}$ and the coefficients of $B(w)$ are fixed by:
\beq
B(w_{i})^{2} \left(  {\alpha}w_{i}^2 + {\beta}w_{i} + {\gamma} \right) = {\nu}_{i}^{2} \, , \qquad
i = 1, \ldots , n
\label{eq:br}
\eeq
Of course, solving the Eqs. \eqref{eq:br} explicitly is not possible. But the connection to gauge theory gives us a good idea as to what these solutions are. For example, in the $n=4$ case the corresponding theory is $SU(2)$ with $N_f = 4$ flavors. So the six solutions
to \eqref{eq:br} in this case are the vacua where one of the quarks, or a monopole, or a dyon becomes massless. The analogous analysis in the $n > 4$ case is simple using the correspondence described in detail in \cite{G}. 

\subsection{Black holes}

Consider the free motion of a probe particle in the background of 
Schwarzschild black hole of mass $M$ (we set $G{N}=1$):
\begin{multline}
ds^2 = \left( 1 - \frac{2M}{r} \right) dt^2 -  \left( 1 - \frac{2M}{r} \right)^{-1} dr^2 - r^2 d{\Omega}_{2}^{2} = \\
\frac{32 M^{3}e^{-\frac{r}{2M}}}{r} du_{+}du_{-} - r^2 d{\Omega}_{2}^{2} 
\end{multline}
where
\beq
u_{\pm} = \pm \sqrt{\frac{r}{2M}-1} \ {\exp} \, {\frac{r \pm t}{4M}}
\eeq
are the Kruskal-Szekeres coordinates, expressed through the Schwarzschild
asymptotic time $t$ and the radial variable $r$. The line element on the two-sphere
$d{\Omega}_{2}^{2} = d{\theta}^{2} + {\rm sin}^2({\theta})d{\phi}^{2}$. 
The geodesic equation, written in the Hamilton-Jacobi form, reads as:
\beq
\left( \frac{r}{2M} \right)^{3} \, e^{r\over 2M} {\partial}_{u_{+}} S {\partial}_{u_{-}}S - 
4\left( \left( {\partial}_{\theta}S \right)^{2} + \frac{1}{{\rm sin}^{2}({\theta})} \left( {\partial}_{\phi}S \right)^{2} \right) = (2 m r)^2
\eeq
for the probe mass $m$. Separating the variables via:
\beq
S (u_{+}, u_{-} ; {\theta}, {\phi} ) = n {\phi} +  {\sl s} ({\theta}) + E\, {\rm log}(u_{+}/u_{-}) +
{\Sigma} ( {\rm log}(u_{+}u_{-}) ) 
\eeq
we obtain, for the radial generating function $\Sigma$:
\beq
{\Sigma}(x) = 4 M p \, \int^{x} w \frac{dz}{z(z-1)} \, , \eeq
with 
\beq 
  w^2   = z ( z^3 + {\mu}^{2} z^{2} + {\nu}^{2} (1-z) )  
  \label{eq:ellcsch}
\eeq
where $z = r/2M$,  ${\mu} = m/p$, ${\nu} = L/(2Mp)$, $p^2 = {\ep}^{2} - m^{2}$, ${\ep} = E/(4 M)$, $E$ is the energy measured by the outside observer, $L$ is the angular momentum\footnote{The angular potential ${\sl s}({\theta})$ is given by 
\beq
{\sl s}({\theta}) = \int \sqrt{L^{2} {\rm sin}^{2}({\theta})- n^2} \frac{d{\theta}}{{\rm sin}({\theta})} \ ,
\eeq and its explicit form is not important for us} and
\beq
x = {\rm log}(u_{+}u_{-}) = z + {\rm log}\left( z- 1 \right) \, , \qquad
dx = \frac{dz}{(1-z^{-1})}
\eeq
is the tortoise coordinate. 

We observe that \eqref{eq:ellcsch} is, again, an elliptic curve, when the variables $p$ and $z$ are extended to the complex domain. We thus expect that the tunneling under the horizon picture \cite{Parikh} of the Hawking radiation \cite{Hawk} should be properly formulated in terms of the periodic orbits on such a curve, as in the previous quantum mechanical examples. It would be interesting to explore this issue further, also in the context of charged and rotating black holes.

\section{Summary and future directions}

This paper is a result of an attempt to write an introduction to the book on modern instanton 
calculations. We found true critical points of 
the classical mechanical actions in a variety of examples. Even though these examples are special in that the complexification of the classical system is an algebraic integrable system, 
we believe the qualitative picture of these solutions can be used in more general problems. 
One qualitative feature of our solutions is that they are the non-linear superpositions
of the instantons, antiinstantons, and perturbative quanta. The former correspond to the $B$-cycle wrappings, while the latter correspond to the $A$-cycle wrappings. 

\subsection{Tropical instantons}

The algebraic integrability matches the perturbative modes of the system to the tunneling modes. The instanton-antiinstanton configurations then arise in the tropical limit of the 
algebraic integrable system, where the underlying spectral curve degenerates to the rational one, while its Jacobian becomes a combinatorial object.

\subsection{Lefschetz thimbles and Dyson-Schwinger equations}

The analytic properties of quantum mechanical correlators can be 
understood by exploring the Lefschetz thimbles of the complexified action
functional. Imagine the correlation functions obey some differential-difference equations in the coupling constants, such as the Dyson-Schwinger equations\footnote{Unfortunately the DS equations do not form a closed system unless one introduces the couplings for all irrelevant operators, or takes some limit}, which come from the invariance of the integral under the small perturbations of the contour of integration. Then the same equation is obeyed by the integrals over all Lefschetz thimbles. The order of such equation is equal to the number of those thimbles. Of course, it is infinite. But it is interesting how infinite it is. It appears that in the system with $r$ degrees of freedom one gets a cone in a lattice of rank $2r$. Each point in this cone corresponds to a solution of the system of those Dyson-Schwinger equations.

Already for one degree of freedom we found many more solutions than we could find in the literature. It appears that the bions of  \cite{Argyres:2012ka, Basar:2013eka, Basar:2015xna} are the particular examples of our solutions for special values of $(m,n)$. 

We found that the critical points of the action functional on the complexified loop space
are the critical points of a flux-type superpotential defined on the (cover of the) base
of an algebraic integrable system. It would be interesting to study Lefschetz thimbles of this superpotential, and compare them to the infinite-dimensional Lefschetz thimbles of the path integral. 

\subsection{Topological renormalisation group and hyperk\"ahler metrics}

Explicit description of the topological renormalisation group flow may prove very difficult. However, 
we have a lot of freedom in the choice of the metric on the space of fields. For example,
one may choose a metric on the space of loops of $L{\CalX}^{2r}_{\BC}$ 
which is induced from the hyperk\"ahler metric on ${\CalX}^{2r}_{\BC}$. The latter, in turn, 
can be simplified by taking a limit of small K\"ahler class of the fibers, the semi-flat metric
discussed in \cite{GMN}. This metric is singular, but the non-singular nearly semi-flat one might work as well.  In this way we'll get the tropical limit of the Lefschetz timble. 

It would be interesting to understand the connection of our solutions to the 
BPS states of supersymmetric gauge theory whose moduli space of vacua is the complexified phase space (it is the four dimensional theory compactified on a circle). 

\subsection{Rademacher for Gutzwiller}

It is tempting to speculate that Gutzwiller's formula \cite{Gutzwiller} can be improved (a la Rademacher) by the contributions of our ${\bf m,n}$-solutions, so as 
to produce an exact formula.  What are the implications for the theory of
quantum chaos?

\subsection{Towards quantum field theory: finite-gap vs lattice}

Of course, the main goal of this project is to improve on the instanton gas ansatz in quantum
field theory, as it fails already in two dimensions \cite{Pol1}. In going from quantum mechanics to quantum field theory one may approximate
the infinite-dimensional system by a finite-dimensional one. The naive approach where we
put the fields on a lattice, may prove too difficult, as all the symmetries would be lost. 
It would be great to find a finite-dimensional approximation which preserves, say, an algebraic integrability of the complexified model. Of course, for some models, such as the sine-Gordon in two dimensions, one can find an integrable lattice approximation \cite{FBA, Jimbo:2008kn}, but such examples are rather exceptional. However, one can find other
finite-dimensional approximations, namely the finite-gap subspaces. It appears that such an approach works for the $O(n)$-sigma models in two dimensions \cite{NK}. In fact, it works better for $n>3$ where the proper two dimensional instantons are simply absent, since ${\pi}_{2}(S^{n-1}) = 0$ in this case!

Recall that the $O(3)$-model can be mapped to the sine-Gordon theory using Pohlmeyer reduction. Since the sine-Gordon equation is an infinite-dimensional integrable system, whose classical (complexified) evolution  linearizes on the Jacobian of an infinite-dimensional complex curve, one could jump at the conclusion that the analogue of our $\bf m, n$-solutions is easy to produce: simply look for the rational windings on that Jacobian
variety. Unfortunately Pohlmeyer reduction works simply only in the infinite volume systems, while our systematic approach requires taking both space and time compact. By imposing periodic or twisted periodic boundary conditions on the sigma model side we arrive at a complicated problem on the sine-Gordon side (the infinite genus curve approach is simple when it is the sine-Gordon equation which is studied with periodic spatial 
boundary conditions). Fortunately, there is another approach, which uses the linear sigma model realization of the $O(n)$ model (which is usually studied in the large $n$ limit), in which one directly gets the solutions of the complexified sigma model, in terms of analytic curves, called the Fermi curves in \cite{NK}. The finite-gap approximation is then the ansatz in which the curve in question has a finite genus component. The motion linearizes on the Prymian of that curve \cite{NK}. In fact, even more explicit solutions can be found using the ``winding ansatz'', in which the spatial slice of the worldsheet winds along an orbit of a $U(1)$-subgroup of $O(n)$ (or, for the twisted boundary conditions, one may choose a generic one-parametric subgroup). In this way the $1+1$ dimensional problem 
reduces to the $0+1$ dimensional one: that of Neumann system, which, in turn, can be reduced to the Gaudin model we discussed above \cite{NT}. Remarkably, despite the common lore that the ${\BC\BP}^{n}$ model is not integrable, a (smaller) class of similar complex solutions can be found there as well.

Encouraged by the success in two dimensions we may hope to find 
the analogous solutions in the four dimensional (non-supersymmetric) 
Yang-Mills theory. The instanton methods in QCD have a very long history \cite{Callan:1977gz}. However, they dilute instanton gas is not a good approximation to the Yang-Mills path integral. We should, therefore, look for the analogues of our non-linear superpositions
of the instantons, antiinstantons, and perturbative quanta (gluons). 
As an initial approach, again, we take the Euclidean space-time with the 
geometry $S^{3} \times S^{1}$. Let $R$ be the radius of $S^3$ and $\beta$ the circumference of $S^1$. We then look for solutions of the Yang-Mills equations
\beq
D_{A}\star F_{A} = 0
\label{eq:ymfe}
\eeq
with the complex gauge field $A$ (say ${\mathfrak{sl}}(2,{\BC})$-valued for the $SU(2)$ gauge theory). The equations \eqref{eq:ymfe} are still conformally invariant, so we may study them on ${\BR}^{4} \backslash 0$ then impose the invariance with respect to the
discrete scaling $r \mapsto r e^{-{\beta}/R}$. The equations on ${\BR}^{4} \backslash 0$
can be studied as in \cite{W77} by imposing an $SO(3)$-invariance on the gauge fields (with $SO(3)$ acting both as the space rotations $SO(3) \subset SO(4)$ and as the gauge rotations). In this way we arrive at the
abelian Higgs model on the periodic $AdS_{2}$, i.e. 
the strip ${\tau}$, ${\rm Im}({\tau})> 0$, ${\tau} \sim {\tau}+1$, 
with the metric
\beq
\frac{d{\tau}d{\bar\tau}}{{\rm Im}({\tau})^2}
\eeq
Perhaps this is not far from the sigma models which can be analyzed using the Fermi curve approach of \cite{NK}. Hopefully in this way the instanton liquid picture of \cite{SS} will be
eventually justified. With more symmetry the $3+1$ dimensional problem reduces to the $0+1$ dimensional one, in fact, again to the anharmonic oscillator \cite{NT}.

\subsection{Wilder speculations}

Note that in order to get an insight on the quantum mechanical model we lifted it to the two dimensional theory and then to the  four dimensional supersymmetric gauge theory in a special background. If we naively adopt the $d \mapsto d+1+2$ point of view 
we would conclude that the non-perturbative treatment of a non-supersymmetric four dimensional quantum field theory should reveal a seven dimensional
supersymmetric theory? There are many supersymmetric theories in seven dimensions: essentially all of them are given by compactifications of M-theory on four-manifolds possibly with fluxes (in the absence of fluxes the four manifolds are hyperk{\"a}hler). Of course, there are many local K3's but probably not as many as four dimensional quantum field theories. Is the existence of the seven dimensional completion another constraint\cite{VS} on the landscape of consistent four dimensional quantum field theories?

\end{document}